\documentclass{article}

\usepackage{graphicx}
\usepackage[normalem]{ulem}
\usepackage{graphicx}
\usepackage{amssymb}
\usepackage{color}
\usepackage{amsmath}
\usepackage{empheq}
\usepackage[capitalize,nameinlink]{cleveref}[0.19]
\usepackage[scale=0.75]{geometry}

\Crefname{figure}{Figure}{Figures}

\crefformat{equation}{\textup{#2(#1)#3}}
\crefrangeformat{equation}{\textup{#3(#1)#4--#5(#2)#6}}
\crefmultiformat{equation}{\textup{#2(#1)#3}}{ and \textup{#2(#1)#3}}
{, \textup{#2(#1)#3}}{, and \textup{#2(#1)#3}}
\crefrangemultiformat{equation}{\textup{#3(#1)#4--#5(#2)#6}}%
{ and \textup{#3(#1)#4--#5(#2)#6}}{, \textup{#3(#1)#4--#5(#2)#6}}{, and \textup{#3(#1)#4--#5(#2)#6}}

\Crefformat{equation}{#2Equation~\textup{(#1)}#3}
\Crefrangeformat{equation}{Equations~\textup{#3(#1)#4--#5(#2)#6}}
\Crefmultiformat{equation}{Equations~\textup{#2(#1)#3}}{ and \textup{#2(#1)#3}}
{, \textup{#2(#1)#3}}{, and \textup{#2(#1)#3}}
\Crefrangemultiformat{equation}{Equations~\textup{#3(#1)#4--#5(#2)#6}}%
{ and \textup{#3(#1)#4--#5(#2)#6}}{, \textup{#3(#1)#4--#5(#2)#6}}{, and \textup{#3(#1)#4--#5(#2)#6}}

\title{Sensitivity Analysis of Chaotic Systems \\ using Unstable Periodic Orbits\footnote{Published version: SIAM Journal on Applied Dynamical Systems, 17(1), 2018 }} 
\author{Davide Lasagna}
\date{}

\begin{document}

\maketitle

\begin{abstract}
A well-behaved adjoint sensitivity technique for chaotic dynamical systems is presented. The method arises from the specialisation of established variational techniques to the unstable periodic orbits of the system. On such trajectories, the adjoint problem becomes a time periodic boundary value problem. The adjoint solution remains bounded in time and does not exhibit the typical unbounded exponential growth observed using traditional methods over unstable non-periodic trajectories (Lea et al., Tellus 52 (2000)). This enables the sensitivity of period averaged quantities to be calculated \emph{exactly}, regardless of the orbit period, because the stability of the tangent dynamics is decoupled effectively from the sensitivity calculations. We demonstrate the method on two prototypical systems, the Lorenz equations at standard parameters and the Kuramoto-Sivashinky equation, a one-dimensional partial differential equation with chaotic behaviour. We report a statistical analysis of the sensitivity of these two systems based on databases of unstable periodic orbits of size $\sim 10^5$ and $\sim 4\times 10^4$, respectively. The empirical observation is that most orbits predict approximately the same sensitivity. The effects of symmetries, bifurcations and intermittency are discussed and future work is outlined in the conclusions.
\end{abstract}


\section{Introduction}

Sensitivity analysis refers broadly to mathematical techniques that, applied to a given system, provide the gradient  of an output quantity of interest with respect to   design  variables parametrising the system at hand. There is a large range of problems across the physical and engineering sciences where sensitivity analysis techniques are tremendously useful: gradient-based optimisation, inverse problems, control design, model calibration, analysis and reduction or uncertainty quantification. 

For time dependent problems, it is common for the output quantity to be the time average of an observable over a finite reference trajectory, generated by a parametrised dynamical system as the solution of an initial value problem from some given initial condition. The interest is typically on the asymptotic state of this trajectory and how   parameters affect the  time average of the observable. Traditional sensitivity techniques, both `tangent' and `adjoint' methods, attempt to calculate the \emph{sensitivity of the time average} with respect to the parameters, the sensitivity of {the} asymptotic state, by calculating the \emph{average of the sensitivity}, i.e. by time averaging the solution of a linear initial value problem obtained from linearisation of the dynamics around {the} reference trajectory \cite{Thuburn:2005ez}.

For chaotic dynamical systems, however, the initial value problem determining the evolution of a trajectory from a given initial condition is inherently ill-conditioned \cite{Wang:2013cx}. Small perturbations to the system parameters have exponentially growing effects on the \emph{future} evolution of unstable trajectories. In such settings, the instability of the tangent dynamics is inextricably intertwined to the sensitivity calculations. Linearised equations, both tangent and adjoint, do not reproduce the inevitable saturation of exponentially growing modes that will occur later on at some point in time. The resulting \emph{average of the sensitivity} leads to unphysically large gradients and does not coincide in the limit with the required \emph{sensitivity of the time average}  \cite{LEA:2000cr, Eyink:2004gk, Thuburn:2005ez, Wang:2013cx}.

In this paper, we aim to show that these well known limitations can be completely overcome if time periodic trajectories, i.e. unstable periodic orbits (UPOs), are used as reference trajectories for sensitivity analysis, rather than using chaotic non-periodic trajectories.
For a time periodic trajectory, it is natural to drop the concept of initial conditions and the causality associated with the forward direction of time. As a results, it is more proficuous to think of UPOs as solutions of a \emph{nonlinear time periodic boundary value problem}, i.e. as trajectories satisfying at each point the governing equations \emph{plus} a periodicity constraint, rather than a simple initial condition. Crucially, this boundary value problem is well-conditioned: excluding pathological cases, i.e. bifurcations, small parameter perturbations have a bounded effect on the periodic solutions, \emph{regardless of their stability characteristics and length}. In other words, small parameter perturbations do not result in exponentially diverging trajectories, but in a bounded state space displacement/distortion of an orbit that remains periodic in this process.   The linearisation of the dynamics around a periodic trajectory   leads to a \emph{linear time periodic boundary problem} that does not admit exponentially growing modes but only a periodic solution. Stability and sensitivity are effectively decoupled and the \emph{average of the sensitivity} coincides with the \emph{sensitivity of the average}, the sensitivity of the asymptotic (periodic) state. This approach shares similar perspectives, discussed in the paper, to those underpinning the Least Squares Shadowing (LSS) method \cite{Wang:2014bt}, recently introduced to circumvent the above mentioned limitations of traditional sensitivity approaches on open trajectories.
 
 
There are also other deeper system theoretical reasons that make UPOs attracting.
For instance, {Periodic Orbit Theory introduces appropriate weighted sums over UPOs to express ergodic statistics of hyperbolic systems \cite{Auerbach:1987baa}, although, it appears that there is no established theory that expresses the sensitivity of the asymptotic state from the sensitivity of UPOs. In this paper, however, we do not aim to deal with fundamental aspects of Periodic Orbit Theory. Rather, our aim is to show that a well-behaved sensitivity technique can be formulated on unstable time periodic trajectories.} The methods that we outline in this paper are analogous to variational methods for time periodic systems previously reported in the literature, for the tangent approach \cite{Vytyaz:2007tk, Wilkins:2009kq} and for the adjoint approach \cite{GIANNETTI:2010jb, Hwang:2014jn, Zahr:2016bd}. However, in the vast majority of cases these methods have been discussed, for a variety of reasons, in the context of problems where the periodic trajectory is \emph{stable}. These formalisms, though, apply regardless of the stability of the orbit, although more care must be taken in the numerical solution.



The main contributions of this paper are thus: the discussion of tangent and adjoint sensitivity techniques for time periodic trajectories and associated numerical algorithms, the discussion of connections to the above-mentioned limitation of traditional initial value approaches and the discussion of potential pitfalls and caveats of the approach. To this end, we demonstrate the method on UPOs of the Lorenz equations at the standard parameters, in \cref{sec:lorenz}, and then consider a model problem relevant to applications, i.e. the sensitivity with respect to feedback control for UPOs of the Kuramoto-Sivashinsky equation, in \cref{sec:kuramoto}. Their modest size makes them convenient playgrounds to test ideas and develop algorithms and enables a statistical analysis of UPO sensitivity, {across inventories of UPOs of size $\sim 10^5$ and $4\times 10^4$, respectively.}

{The sensitivity methods reported in this paper only apply to systems with discrete symmetries. For instance, for the KS equation, we restrict the evolution on the flow-invariant subspace of odd solutions, resulting in a dynamical system with a discrete symmetry, \cite{Lan:2008kg, Lan:2004ch}. Methods for sensitivity analysis of time-periodic invariant solutions of systems with continuous symmetries are left to future work.}

Tangent and adjoint sensitivity techniques for open (non-periodic) trajectories are first presented, in \cref{sec:sensitivity-chaotic-trajectories}. These methods are well know but are reported here to highlight the differences with analogous sensitivity techniques for time periodic trajectories, discussed in \cref{sec:sensitivity-periodic-trajectories}. For convenience, we use a finite-dimensional setting and assume the dynamical system is defined by a set of ordinary differential equations (ODEs). Similar techniques can be also be developed in the continuous setting for partial differential equations.

\section{Tangent and adjoint sensitivity of non-periodic unstable trajectories}\label{sec:sensitivity-chaotic-trajectories}

Let
\begin{subequations}\label{eq:system}
\begin{empheq}[left={\empheqlbrace}]{align}
	\dot{\mathbf{x}}(t) &= \mathbf{f}(\mathbf{x}(t); \mathbf{p}),\quad  t \in [0, T]\\
	\mathbf{x}(0) &= \mathbf{x}_0\label{eq:system-ic}
\end{empheq}
\end{subequations}
be {an} initial value problem (IVP), where $\mathbf{x}(t) \in \mathbb{R}^{N_\mathbf{x}}$ {denotes} the reference state space trajectory originating at some initial condition $\mathbf{x}_0$ and $t$ is time. The field \mbox{$\mathbf{f} : \mathbb{R}^{N_\mathbf{x}} \times \mathbb{R}^{N_\mathbf{p}} \rightarrow \mathbb{R}^{N_\mathbf{x}}$} is a smooth vector function of $\mathbf{x}$ and depends, additionally, on a vector of variables $\mathbf{p} \in \mathbb{R}^{N_\mathbf{p}}$, the design  variables of the problem. The variables $\mathbf{p}$ parametrise implicitly, in some unknown fashion, the trajectories of \cref{eq:system}. Let now
\begin{equation}
K(t) = K(\mathbf{x}(t), \mathbf{p}) : \mathbb{R}^{N_\mathbf{x}}\times\mathbb{R}^{N_\mathbf{p}} \rightarrow \mathbb{R}
\end{equation}
denote an observable of the system,   also referred to as cost in the following. Its   average over the reference trajectory,
\begin{equation}\label{eq:long-time-average}
\displaystyle \overline{K} = \frac{1}{T}\int_0^T K(\mathbf{x}(t), \mathbf{p}) \mathrm{d}t,
\end{equation}
will depend implicitly on the parameters too. 

%
%
 The quantity of interest is the \emph{sensitivity of the average} to {parameter perturbations,} expressed by the gradient $\raisebox{1.5pt}{(}\overline{K}\raisebox{1.5pt}{)}_{\mathrm{d}\mathbf{p}}$. In practice, sensitivity techniques only provide the \emph{average of the sensitivity}, obtained by time averaging {the result} of a linear problem (see discussion in \cite{Thuburn:2005ez}). In the tangent approach, the average of the sensitivity is obtained by differentiating \cref{eq:long-time-average} with respect to the parameters, leading to 
\begin{equation}\label{eq:gradient-of-average}
	{\overline{(K_{\mathrm{d}\mathbf{p}})} = \frac{1}{T}\int_0^T \big[K_{\partial\mathbf{x}}(t) \cdot \mathbf{x}_{\partial \mathbf{p}}(t) + K_{\partial \mathbf{p}}(t)\big]\,\mathrm{d}t,}
\end{equation}
where subscripts `$\mathrm{d}$' and `$\partial$' denote total and partial differentiation. We follow the notation that gradients are row vectors. 

The quantity $\mathbf{x}_{\partial\mathbf{p}}(t) \in \mathbb{R}^{N_{\mathbf{x}}\times N_\mathbf{p}}$ is the state sensitivity. The linear problem for its evolution is obtained by differentiating \cref{eq:system} with respect to the parameters, leading to the   tangent equations, the IVP
\begin{subequations}\label{eq:tangent-sensitivity-open}
\begin{empheq}[left={\empheqlbrace}]{align}
\dot{\mathbf{x}}_{\partial \mathbf{p}}(t) &= \mathbf{f}_{\partial \mathbf{x}}(t) \cdot \mathbf{x}_{\partial \mathbf{p}}(t) + \mathbf{f}_{\partial \mathbf{p}}(t), \quad t \in [0, T]\\
  	\mathbf{x}_{\partial \mathbf{p}}(0) &= \mathbf{0}.\label{eq:tangent-sensitivity-open-ic}
\end{empheq}
\end{subequations}
where $\mathbf{f}_{\partial\mathbf{x}}(t) \in \mathbb{R}^{N_\mathbf{x} \times N_\mathbf{x}}$ and $\mathbf{f}_{\partial\mathbf{p}}(t) \in \mathbb{R}^{N_\mathbf{x} \times N_\mathbf{p}}$, the Jacobians of the vector field, contain the partial derivatives of $\mathbf{f}$ with respect to the state and parameters, respectively.
The state sensitivity describes the effects that parameter perturbations have on the \emph{future} evolution of state space trajectories originating from the same initial condition $\mathbf{x}_0$, hence $\mathbf{x}_{\partial \mathbf{p}}(0) = \mathbf{0}$. For an unstable reference trajectory, the state sensitivity grows {exponentially at an average rate} determined by the leading Lyapunov exponent of the system. This is a reflection of the ill-conditioning of the IVP \cref{eq:system}, whereby small parameter perturbations lead to large changes in the solution \cite{Wang:2014hu}. When the length $T$ of the trajectory is increased, with the idea of obtaining the sensitivity of converged averages, the integral \cref{eq:gradient-of-average} does not converge to a finite value but grows exponentially and yields unphysical gradients \cite{LEA:2000cr,Thuburn:2005ez,Wang:2013cx}.


The same limitation applies when an adjoint approach is used to obtain the gradient \cref{eq:gradient-of-average}, even if the state sensitivity $\mathbf{x}_{\partial \mathbf{p}}(t)$ is never computed explicitly. 
The governing equation \cref{eq:system} is adjoined to the time average \cref{eq:long-time-average}, via the vector of adjoint variables $\boldsymbol \lambda(t) \in \mathbb{R}^{N_\mathbf{x}}$, forming the Lagrangian
\begin{equation}\label{eq:lagrangian-initial-open}
	\mathcal{L} = \frac{1}{T}\int_0^{T} K(t) + \boldsymbol \lambda^\top(t)\cdot\big[\dot{\mathbf{x}}(t) - \mathbf{f}(t)\big]\,\mathrm{d}t.
\end{equation}

Integrating by parts with respect to time the term $\boldsymbol \lambda^\top\cdot\dot{\mathbf{x}}$ leads to
\begin{equation}\label{eq:lagrangian}
	\displaystyle \mathcal{L} =  \frac{1}{T}\int_0^{T} K(t) - \dot{\boldsymbol \lambda}^\top(t) \cdot\mathbf{x}(t) - \boldsymbol \lambda^\top(t)\cdot \mathbf{f}(t) \,\mathrm{d}t \; +\;  \big[\boldsymbol \lambda^\top(t)\cdot\mathbf{x}(t)\big]_{0}^{T},
\end{equation}
and differentiation with respect to the parameters yields
\begin{align}\label{eq:lagrangian-final-open}
 	\mathcal{L}_{\mathrm{d}\mathbf{p}} = \overline{(K_{\mathrm{d}\mathbf{p}})} =  \frac{1}{T}\int_0^{T}  &\underbrace{\big[ K_{\partial\mathbf{x}}(t) - \dot{\boldsymbol \lambda}^\top(t) - \boldsymbol \lambda^\top (t)\cdot \mathbf{f}_{\partial\mathbf{x}}(t)\big]}_\mathbf{a}\cdot \mathbf{x}_{\partial\mathbf{p}}(t) - \boldsymbol \lambda^\top(t)\cdot\mathbf{f}_{\partial\mathbf{p}}(t) + K_{\partial \mathbf{p}}(t) \,\mathrm{d}t \;+ \nonumber \\ + & \underbrace{\big [ \boldsymbol \lambda^\top(t) \cdot \mathbf{x}_{\partial\mathbf{p}}(t)\big]_{0}^{T}}_\mathbf{b},
\end{align}
where the first equality holds because it is assumed that when parameters are perturbed the governing equation remains satisfied, for any choice of $\boldsymbol \lambda(t)$. 
The adjoint variables are selected such that $\mathbf{a}$ and $\mathbf{b}$ vanish, circumventing the (expensive) calculation of the state sensitivity. Requiring $\mathbf{a}=\mathbf{0}$ yields the adjoint equation \cref{eq:adjoint-problem-open-eq}, while its initial conditions are obtained by requiring $\mathbf{b}=\mathbf{0}$, i.e.
\begin{equation}
	\mathbf{0} = \mathbf{b} = \big [ \boldsymbol \lambda^\top(t) \cdot \mathbf{x}_{\partial\mathbf{p}}(t)\big]_{0}^{T} = \boldsymbol \lambda^\top(T) \cdot \mathbf{x}_{\partial\mathbf{p}}(T) - \boldsymbol \lambda^\top(0) \cdot \mathbf{x}_{\partial\mathbf{p}}(0)
\end{equation}
For any $\boldsymbol \lambda (0)$, the second term at $t=0$ is zero because the initial conditions are assumed not to change, while for any $\mathbf{x}_{\partial \mathbf{p}}(T)$, the first term at $t=T$ vanishes if $\boldsymbol \lambda(T)=0$. This results in the adjoint problem 
\begin{subequations}\label{eq:adjoint-problem-open}
\begin{empheq}[left={\empheqlbrace}]{align}
	\dot{\boldsymbol \lambda}(t) &= -\mathbf{f}^\top_{\partial\mathbf{x}}(t)\cdot \boldsymbol \lambda (t) + K^\top_{\partial\mathbf{x}}(t)\label{eq:adjoint-problem-open-eq}\\
	\boldsymbol \lambda(T) &= \mathbf{0}\label{eq:adjoint-problem-open-bc}, 
\end{empheq}
\end{subequations}
a linear initial value problem with homogeneous terminal conditions. {Because of such conditions,} problem \cref{eq:adjoint-problem-open} needs to be solved \emph{backward} in time. This adds a number of well-known complications since the time varying coefficients in \cref{eq:adjoint-problem-open} are defined on a trajectory that proceeds \emph{forward} in time.

Similarly to the tangent sensitivity problem, the solution of \cref{eq:adjoint-problem-open} grows at an average exponential rate when integrated in time over an unstable trajectory. Hence, the integral
\begin{equation}\label{eq:adjoint-gradient-open}
	\mathcal{L}_{\mathrm{d}\mathbf{p}} = \overline{(K_{\mathrm{d}\mathbf{p}})} =  \frac{1}{T}\int_0^{T}  - \boldsymbol \lambda^\top(t)\cdot\mathbf{f}_{\partial\mathbf{p}}(t) + K_{\partial \mathbf{p}}(t)  \,\mathrm{d}t.
\end{equation}
does not converge as the length of the trajectory is increased, but the gradient gets exponentially larger, and unphysical, for increasing $T$ \cite{LEA:2000cr,Eyink:2004gk,Thuburn:2005ez,Wang:2013cx}. 


\section{Tangent and adjoint sensitivity of periodic unstable trajectories}\label{sec:sensitivity-periodic-trajectories}
  For time periodic problems, it is practical to first introduce the so-called Linsteadt-Poincar\'e transformation 
\begin{equation}
s = 2\pi t/T = \omega t,
\end{equation}
with  $\omega$ being the fundamental frequency of a particular reference periodic trajectory $\mathbf{x}^o(s)$, to uniformly transform time $t\in [0, T]$ into the loop parameter $s\in [0, 2\pi]$. This transformation makes the natural frequency appear explicitly in the evolution equation
\begin{subequations}\label{eq:loop-equation}
\begin{empheq}[left={\empheqlbrace}]{align}
	\omega \dot{\mathbf{x}}^o(s) &= \mathbf{f}(\mathbf{x}^o(s); \mathbf{p}),\quad s\in [0, 2\pi]\\
	\mathbf{x}^o(0) &= \mathbf{x}^o(2\pi)
\end{empheq}
\end{subequations}
and removes the explicit dependence of the period $T$ from the limits of integration in the calculation of the period averages, denoted by brackets, i.e.
\begin{equation}\label{eq:orbit-average} 
	\langle K \rangle = \frac{1}{2\pi}\int_0^{2\pi} K(\mathbf{x}^o(s), \mathbf{p})\,\mathrm{d}s.
\end{equation}

Problem \cref{eq:loop-equation} is a nonlinear time periodic boundary value problem (BVP). Its solutions, infinitely many for a chaotic system, are pairs $\{\mathbf{x}^o(s), \omega\}$, pairs of unstable periodic orbits and the associated {frequencies}. Numerical methods to solve BVPs like \cref{eq:loop-equation} are well known \cite{Ascher:1994ty}. In this paper, the modest size of the systems we consider enables us to use an approach known as \emph{quasi-linearisation} in the literature, to reduce the nonlinear problem to a sequence of linear BVPs solved with standard finite difference technique for temporal discretisation (see \cref{app:upo-search-algo,app:appendix-numerical-solution}). {For high-dimensional systems, shooting strategies might be more practical \cite{Ascher:1994ty,Zahr:2016bd}.}

Similarly to solutions of the IVP \cref{eq:system}, the solutions of the BVP (both the trajectory and the frequency) depend implicitly on the parameters $\mathbf{p}$, in some unknown fashion, because the vector field where UPOs reside is parametrised.   {Small perturbations of the BVP \cref{eq:loop-equation}, e.g. a small parameter perturbation}, produces, in general, small changes to its solutions, i.e. a small displacement/distortion of the trajectory and a small change in the frequency. In other words, the condition number of problem \cref{eq:loop-equation} is only `moderate' and does not increase exponentially with the length of the orbit, as for the IVP approach \cite{Wang:2014hu}.

For tangent sensitivity analysis, the gradient of the period average is obtained by differentiating \cref{eq:orbit-average}, leading to
\begin{equation}\label{eq:upo-gradient}
	\langle K_{\mathrm{d}\mathbf{p}} \rangle = \langle K_{\partial\mathbf{x}}(s) \cdot \mathbf{x}_{\partial\mathbf{p}}(s) + K_{\partial\mathbf{p}}(s)\rangle.
\end{equation}
The evolution equation for the state sensitivity $\mathbf{x}_{\partial\mathbf{p}}(s)$ is obtained by differentiating \cref{eq:loop-equation} with respect to the parameters, to obtain
\begin{subequations}\label{eq:tangent-sensitivity-upo}
\begin{empheq}[left={\empheqlbrace}]{align}
	\omega \dot{\mathbf{x}}_{\partial\mathbf{p}}(s) &= \mathbf{f}_{\partial\mathbf{x}}(s)\cdot \mathbf{x}_{\partial\mathbf{p}}(s) + \mathbf{f}_{\partial\mathbf{p}}(s) - \omega_{\partial\mathbf{p}}\dot{\mathbf{x}}^o(s),\quad s\in [0, 2\pi]\label{eq:tangent-sensitivity-upo-eq}\\
	\mathbf{x}_{\partial\mathbf{p}}(0) &= \mathbf{x}_{\partial\mathbf{p}}(2\pi)\label{eq:tangent-sensitivity-upo-bc}\\
	\mathbf{0} &= \mathbf{x}_{\partial\mathbf{p}}(0)^\top\cdot\mathbf{f}(0)\label{eq:tangent-sensitivity-upo-plc}.
\end{empheq}
\end{subequations}
a linear time periodic BVP. Such boundary conditions reflect the fact that the arbitrary orbit origin $\mathbf{x}^o(0) = \mathbf{x}^o(2\pi)$ is a single point that moves in a single direction when parameters are varied, so that the perturbed orbit remains periodic. Unlike in \cref{eq:tangent-sensitivity-open-ic}, $\mathbf{x}_{\partial \mathbf{p}}(0) \neq \mathbf{0}$ and every point on the orbit is allowed to move in state space when parameters are varied. The gradient $\omega_{\partial \mathbf{p}}$ is identically zero for forced oscillations, but it is non-zero and unknown for limit-cycle-type oscillations and its value can be obtained as part of the solution of \cref{eq:tangent-sensitivity-upo}. The phase locking constraint \cref{eq:tangent-sensitivity-upo-plc} ensures that \cref{eq:tangent-sensitivity-upo} has a unique solution. Without this condition, problem \cref{eq:tangent-sensitivity-upo} has a one-parameter family of solutions, corresponding to arbitrary shifts along the orbit: if $\mathbf{x}_{\partial\mathbf{p}}(s)$ is a solution, $\mathbf{x}_{\partial\mathbf{p}}(s+\Delta)$ is a solution too, for any real $\Delta$. Other equivalent forms of this constraint can be also introduced \cite{Lan:2004ch, Wilkins:2009kq}.  

Solutions of \cref{eq:tangent-sensitivity-upo} are time periodic and hence bounded, regardless of the stability characteristics and length of the underlying reference trajectory. The exponential growth of solutions of \cref{eq:tangent-sensitivity-open} over unstable non-periodic trajectories is avoided in a natural fashion by the  choice of the boundary conditions. This implies that the linearisation underpinning \cref{eq:upo-gradient} and \cref{eq:tangent-sensitivity-upo} remains valid regardless of the length or the orbit, and thus
\begin{equation}
	\langle K\rangle_{\mathrm{d}\mathbf{p}} = \langle K_{\mathrm{d}\mathbf{p}} \rangle.
\end{equation}
In other words, the   sensitivity of the asymptotic state (now periodic), i.e. the \emph{sensitivity of the average}, coincides with the \emph{average of the sensitivity} obtained from the linear problem.

As anticipated in the introduction, the proposed sensitivity method for periodic trajectories shares several similarities with Least-Squares Shadowing (LSS) methods for sensitivity analysis of chaotic trajectories \cite{Blonigan:2014je}. The common ground is that the initial condition of the perturbed trajectory is permitted to change, to avoid the exponential divergence. In LSS, the perturbation of the initial condition $\mathbf{x}_{\partial \mathbf{p}}(0)$ and the entire sensitivity solution $\mathbf{x}_{\partial \mathbf{p}}(t)$ are obtained from the solution of a constrained optimisation problem, whereby the average norm of $\mathbf{x}_{\partial \mathbf{p}}(t)$ is minimised subject to the constraint of the sensitivity equation  (Equation 13 in \cite{Wang:2014hu}). In stark contrast, for a periodic trajectory, the sensitivity solution is obtained, more naturally, by requiring temporal periodicity. It is important to point out, though, that for a periodic trajectory, the computed sensitivity is exact, and the method does not display the ``shadowing error'' of LSS \cite{Wang:2014bt} at the trajectory end points.

For the adjoint method, we proceed as in \cref{sec:sensitivity-chaotic-trajectories}, by formulating a Lagrangian and differentiating it with respect to the parameters, to obtain
\begin{multline}\label{eq:lagrangian-final}
 	\mathcal{L}_{\mathrm{d}\mathbf{p}} = \langle K \rangle_{\mathrm{d}\mathbf{p}} = \Big\langle \underbrace{\big[ K_{\partial\mathbf{x}}(s) - \omega \dot{\boldsymbol \lambda}^\top(s) - \boldsymbol \lambda^\top (s)\cdot \mathbf{f}_{\partial\mathbf{x}}(s)\big]}_\mathbf{a}\cdot \mathbf{x}_{\partial\mathbf{p}}(s) - \boldsymbol \lambda^\top(s)\cdot\mathbf{f}_{\partial\mathbf{p}}(s)  + \\ + K_{\partial\mathbf{p}}(s) + \omega_{\partial\mathbf{p}} \boldsymbol \lambda^\top(s)\cdot\dot{\mathbf{x}}^o(s) \Big\rangle  + \underbrace{\big [ \boldsymbol \lambda^\top(s) \cdot \mathbf{x}_{\partial\mathbf{p}}(s)\big]_{0}^{2\pi}}_\mathbf{b}
\end{multline}
which differs from \cref{eq:lagrangian-final-open} only by the last term in the brackets arising from the time transformation. Requiring $\mathbf{a}$ to vanish leads to the adjoint equation \cref{eq:adjoint-equation}, the same as \cref{eq:adjoint-problem-open-eq}. Requiring $\mathbf{b}$ to vanish, i.e.
\begin{equation}\label{eq:condition-adjoint-upo}
	\mathbf{0} = \mathbf{b} = \big [ \boldsymbol \lambda^\top(s) \cdot \mathbf{x}_{\partial\mathbf{p}}(s)\big]_{0}^{2\pi} = \boldsymbol \lambda^\top(2\pi) \cdot \mathbf{x}_{\partial\mathbf{p}}(2\pi) - \boldsymbol \lambda^\top(0) \cdot \mathbf{x}_{\partial\mathbf{p}}(0) = [\boldsymbol \lambda(2\pi) - \boldsymbol \lambda(0)]^\top\cdot\mathbf{c},
\end{equation} 
leads to time periodic boundary conditions \cref{eq:adjoint-bc}, because \mbox{$\mathbf{x}_{\partial \mathbf{p}}(0) = \mathbf{x}_{\partial\mathbf{p}}(2\pi) = \mathbf{c}$} for some unknown $\mathbf{c}$. This results in the adjoint problem
\begin{subequations}\label{eq:adjoint-problem}
\begin{empheq}[left={\empheqlbrace}]{align}
	\omega \dot{\boldsymbol \lambda}(s) &= -\mathbf{f}^\top_{\partial\mathbf{x}}(s)\cdot \boldsymbol \lambda (s) + K^\top_{\partial\mathbf{x}}(s)\label{eq:adjoint-equation}\\
	\boldsymbol \lambda(0) &= \boldsymbol \lambda(2\pi)\label{eq:adjoint-bc}\\
	0 &= \boldsymbol \lambda^\top(0)\cdot\mathbf{f}(0)\label{eq:adjoint-plc} 
\end{empheq}
\end{subequations}
which should be compared to \cref{eq:adjoint-problem-open}. The constraint \cref{eq:adjoint-plc} is required to break the translational invariance; the solution is not unique but there is a one-parameter family of adjoint solutions. 

Similarly to the tangent problem, the adjoint problem is a linear time periodic BVP, the adjoint solution is periodic and bounded and cannot grow exponentially in time. An appropriate numerical method must be used to enforce the boundary conditions. When shooting techniques are used for the solution \cite{Ascher:1994ty, Zahr:2016bd}, the adjoint equation \emph{does not need to be marched backwards} but the ``forward'' and adjoint equations can be marched forward simultaneously, alleviating well-known numerical issues associated to \cref{eq:adjoint-problem-open}. Numerical methods required to solve \cref{eq:adjoint-problem} and \cref{eq:tangent-sensitivity-upo} are fundamentally similar to those used to search periodic trajectories in the first place, enabling reuse of algorithms and computational tools (see \cref{app:appendix-numerical-solution}).

Upon solution, the gradient of the period average can then be calculated as 
\begin{equation}\label{eq:gradient-adjoint-upo}
	\langle K \rangle_{\mathrm{d}\mathbf{p}} = \Big \langle K_{\partial \mathbf{p}}(s) - \boldsymbol \lambda^\top(s)\cdot\mathbf{f}_{\partial\mathbf{p}}(s)  - \omega_{\partial\mathbf{p}} \boldsymbol \lambda^\top(s)\cdot\dot{\mathbf{x}}^o(s)\Big\rangle.
\end{equation}

The unknown frequency gradient still appears in \cref{eq:gradient-adjoint-upo}. Rather than from the (expensive) solution of the tangent sensitivity problem, this quantity can be obtained at little additional cost using Fredholm's alternative (\cite{hale2009ordinary}, pg. 146). Fredholm's alternative expresses a compatibility constraint for the solution of linear time periodic BVPs. Specifically, problem \cref{eq:tangent-sensitivity-upo} has time-periodic solutions if and only if
\begin{equation}
	\int_{0}^{2\pi} \mathbf{w}^\top(s) \cdot \big[ \mathbf{f}_{\partial\mathbf{p}}(s) - \omega_{\partial\mathbf{p}}\dot{\mathbf{x}}^o(s) \big]\,\mathrm{d}s = 0 
\end{equation}
for all $2\pi$-periodic solutions of the homogeneous adjoint problem
\begin{subequations}\label{eq:homo-adjoint-problem}
\begin{empheq}[left={\empheqlbrace}]{align}
	\omega \dot{\mathbf{w}}(s) &= -\mathbf{f}^\top_{\partial\mathbf{x}}(s)\cdot \mathbf{w} (s) \label{eq:homo-adjoint-equation}\\
	\mathbf{w}(0) &= \mathbf{w}(2\pi)\label{eq:homo-adjoint-bc}
\end{empheq}
\end{subequations}
A non-trivial solution can be obtained by augmenting \cref{eq:homo-adjoint-problem} with a inhomogeneous phase locking condition $\mathbf{w}(0)^\top\cdot\mathbf{f}(0) = 1$, and part of the computational cost can be amortised by solving it along with \cref{eq:adjoint-problem}. The gradient of the frequency is then
\begin{equation}
\displaystyle\omega_{\partial \mathbf{p}} = \frac{ \displaystyle{\int_{0}^{2\pi}\mathbf{w}^\top(s)\cdot\mathbf{f}_{\partial\mathbf{p}}(s)}\,\mathrm{d}s}{{\displaystyle \int_{0}^{2\pi}\mathbf{w}^\top(s)\cdot\dot{\mathbf{x}}^o(s)}\,\mathrm{d}s}
\end{equation}


\subsection{Non-uniqueness of the state sensitivity}\label{sec:least-squares}

The gradient $\langle K \rangle_{\mathrm{d} \mathbf{p} }$ quantifies the global, average effects of perturbations but does not convey information about how these are distributed during the UPO period. Knowledge of such a distribution could be useful, for instance, to pinpoint instants in time where parameters have the largest influence on the cost. 

Using the tangent sensitivity formulation, and assuming a single scalar parameter $p$, one could recall \cref{eq:upo-gradient} and use the quantity $ K_{\partial\mathbf{x}}(s)\cdot \mathbf{x}_{\partial p }(s) + K_{\partial p }(s)$ for this purpose. This would be incorrect, however, because the sensitivity $\mathbf{x}_{\partial p }(s)$ fails to convey uniquely the desired insight into the orbit deformation, since the orbit origin, and thus the phase locking condition to break the time-translational invariance, is arbitrary. Similar considerations apply for the adjoint variables. In other words, the global change in state space structure of an orbit is unique, but there is a one-parameter family of time-preserving sensitivity trajectories that express this change locally. These can be generated as 
\begin{equation}
\mathbf{x}_{\partial p }^\tau (s) = \mathbf{x}_{\partial p } (s) + \tau \mathbf{f}(s),
\end{equation}
by a shift along the orbit. Note that for any $\tau$,
\begin{equation}\label{eq:perturbed-sensitivity-integrals}
\langle K_{\partial\mathbf{x}}(s)\cdot\mathbf{x}_{\partial p}(s)\rangle = \langle K_{\partial\mathbf{x}}(s)\cdot\mathbf{x}^\tau_{\partial p}(s)\rangle
\end{equation}
holds, as $\langle K_{\partial\mathbf{x}}(s)\cdot\mathbf{f}(s)\rangle$ is the closed path integral of the conservative vector field $K_{\partial\mathbf{x}}$ and it is identically zero. Here we wish to develop a method to express the local change uniquely. 

The phase locking condition \cref{eq:tangent-sensitivity-upo-plc} selects one sensitivity trajectory by enforcing the sensitivity to be orthogonal to the flow at the arbitrary origin. However, on the rest of the orbit the orthogonality is not guaranteed, i.e., $\mathbf{x}_{\partial p}(s) \cdot \mathbf{f}(s) \neq 0$ in general. The idea, inspired by LSS \cite{Wang:2014hu}, is then to calculate \emph{a posteriori} the shift $\tau^*$ that minimises the average squared norm of the sensitivity, the quantity $\tau^* = \arg \underset{\tau}{\min} \langle \| \mathbf{x}_{\partial p}^\tau(s) \|^2\rangle$. A simple variational calculation shows that this shift is given by 
\begin{equation}
	\tau^* = -{\langle \mathbf{x}_{\partial p}^\top(s)\cdot \mathbf{f}(s)\rangle}/{\langle \mathbf{f}^\top(s) \cdot \mathbf{f}(s)\rangle}.
\end{equation}
This shift is unique for a given periodic trajectory and can be used to define and illustrate the local changes to the orbit structure induced by parameter perturbations.
An alternative, similarly inspired by LSS and which we do not pursue here, is to remove the component of the sensitivity tangent to the direction of the flow $\mathbf{x}_{\partial p}^\shortparallel(s)$ and retain the component orthogonal to it, $\mathbf{x}_{\partial p}^\perp(s)$, to obtain a solution that has least possible squared average. This is achieved by a straightforward orthogonalisation,
\begin{equation}
	\mathbf{x}_{\partial p}^\perp(s) = \mathbf{x}_{\partial p}(s) - \mathbf{x}_{\partial p}^\shortparallel(s) = \mathbf{x}_{\partial p}(s) - \mathbf{n}(s)\big ( \mathbf{x}_{\partial p}(s)\cdot\mathbf{n}(s)\big) 
\end{equation}
so that $\mathbf{x}_{\partial p}^\perp(s) \cdot \mathbf{f}(s) = 0, \forall s\in [0, 2\pi]$ holds, where $\mathbf{n}(s) = \mathbf{f}(s)/\|\mathbf{f}(s)\|$ is the unit-norm loop tangent vector. This operation, however, changes the flow of time and \cref{eq:perturbed-sensitivity-integrals} ceases to hold.

\begin{figure}[tbp]
	\centering
	\includegraphics[width=0.86\textwidth]{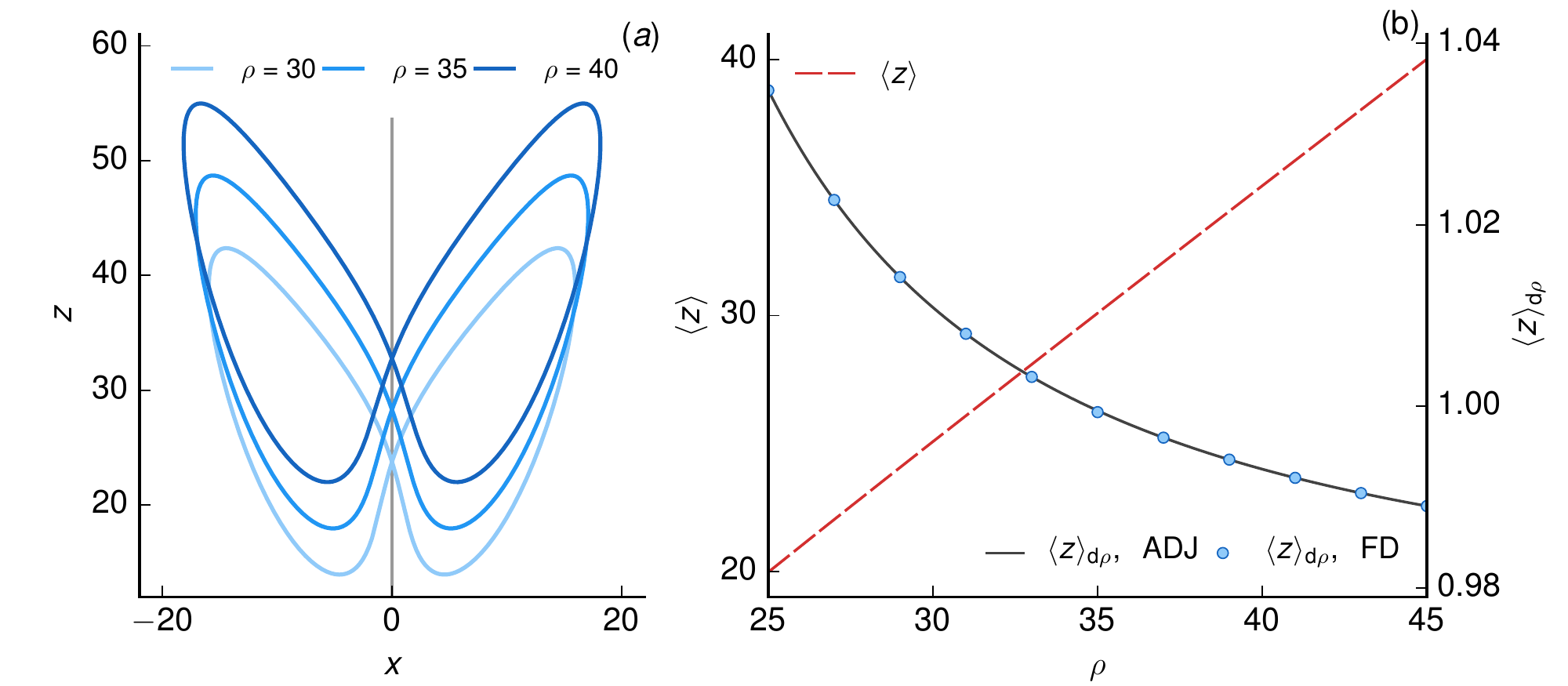}
	\caption{Panel (a): shortest UPO for selected values of $\rho$. Panel (b): variation of {the period average} $\langle z \rangle$ with $\rho$ (dashed line, left axis) and its gradient $\langle z \rangle_{\mathrm{d}\rho}$ (right axis), computed using the adjoint algorithm ($\mathrm{ADJ}$, solid line) and approximated via finite differences ($\mathrm{FD}$, circles).}
	\label{fig:comparison-with-fd-lorenz}
\end{figure}
\section{Application to the Lorenz Equations}\label{sec:lorenz}
The adjoint sensitivity algorithm discussed in \cref{sec:sensitivity-periodic-trajectories} is demonstrated on UPOs of the celebrated Lorenz equations \cite{Lorenz:1963tf}. These are
\begin{eqnarray}\label{eq:lorenz-equations}
  \left\{
  \begin{aligned}
	\dot{x}(t) &= \sigma(x(t)-y(t))\\
	\dot{y}(t) &= \rho x(t) - y(t) - x(t)z(t)\\
	\dot{z}(t) &= x(t)y(t) - \beta z(t)
  \end{aligned}
  \right .
\end{eqnarray}
where the standard parameters $\sigma=10$, $\beta=8/3$ and $\rho = 28$ are used throughout. Let $z$ be the observable of the system, $\rho$ the sensitivity parameter and let us calculate the sensitivity $\langle z \rangle_{\mathrm{d}\rho}$. An approximation of the sensitivity $(\overline{z})_{\mathrm{d}\rho}$ around $\rho=28$ obtained using, e.g., finite difference quotients of long-time averages or similar techniques, is about unity.

We first validate the approach by considering the shortest UPO, with cycle length $n=2$, with $n$ the number of intersections with the standard Poincar\'e section $z=\rho-1$, $\dot{z}(s)>0$. This UPO spans both lobes of the attractor, has symbol sequence AB (in the notation of \cite{Viswanath:2003gb}) and {is invariant under} the symmetry transformation $S : [x, y, z] \rightarrow [-x, -y, z]$. We find numerically this orbit at $\rho=25$ and perform continuation in $\rho$ up to $\rho=45$, with intervals $\Delta \rho = 0.2$. 
UPOs for selected values of $\rho$ are shown in \cref{fig:comparison-with-fd-lorenz}-(a). The {period} average $\langle z \rangle$, shown in panel (b) as a function of $\rho$ (dashed line, left axis), follows approximately a linear behaviour. The slope of this curve, obtained using a second order accurate finite difference approximation for a few values of $\rho$ (blue {circles}, right axis, denoted by FD), spans approximately the range [0.99, 1.04]. The sensitivity $\langle z \rangle_{\mathrm{d}\rho}$ calculated using the adjoint method discussed previously (solid line) matches with the finite difference approximation up to discretisation errors involved in the solution of the relevant BVPs and the truncation/cancellation errors in the finite difference quotient. 


\begin{figure}[htbp]
	\centering
	\includegraphics[width=0.83\textwidth]{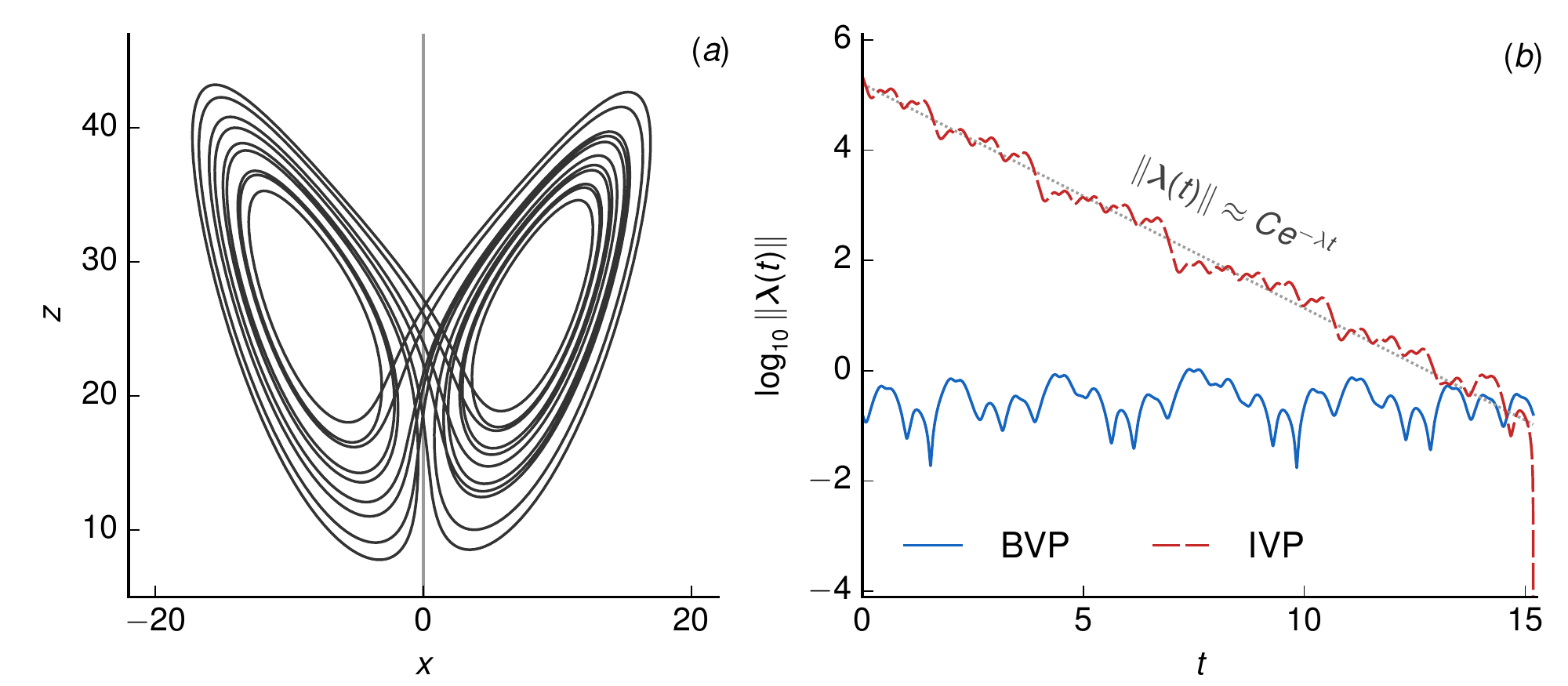}
	\caption{Panel (a): the $n=20$ UPO ($T\approx15.1827$) used as reference trajectory to solve the adjoint BVP \cref{eq:adjoint-problem} and IVP \cref{eq:adjoint-problem-open}. Panel (b): the norm of the adjoint variables from solution of these two problems.}
	\label{fig:classical-adjoint-blowup-norm}
\end{figure}

{We now calculate the sensitivity for a long UPO ($n=20$, \cref{fig:classical-adjoint-blowup-norm}-(a)) using the BVP approach \cref{eq:adjoint-problem}, and the IVP approach \cref{eq:adjoint-problem-open}, using the backward-in-time integration}. For the latter, the UPO is treated as an open trajectory, and we select an arbitrary point {on it} as the initial/final condition. In \cref{fig:classical-adjoint-blowup-norm}-(b), the {time histories} of the norm of the adjoint variables from the solution of two problems are reported. When the traditional IVP approach is used (dashed red line), the norm of the adjoint variables starts from zero, rapidly jumps to a finite value and then grows exponentially at an {average rate equal to the positive Lyapunov exponent of the system at these parameters, $\lambda \approx 0.906$ \cite{Sprott:2003vc}}. The sensitivity of the time average calculated from \cref{eq:adjoint-gradient-open} is unphysical, $(\overline{z})_{\mathrm{d}\rho} \approx 3017.6$. In stark contrast, the norm of the adjoint variables from the solution of the BVP \cref{eq:adjoint-problem} (solid blue line) is a periodic function  . The sensitivity of the period average \cref{eq:gradient-adjoint-upo} is physically meaningful, $\langle z \rangle_{\mathrm{d}\rho} \approx 1.01847$. This applies regardless of the length of the UPOs (see \cref{app:long-upo-lorenz} for sensitivity analysis of an UPO with $T\approx1080.5$). 


\begin{figure}[htbp]
	\centering
	\includegraphics[width=0.95\textwidth]{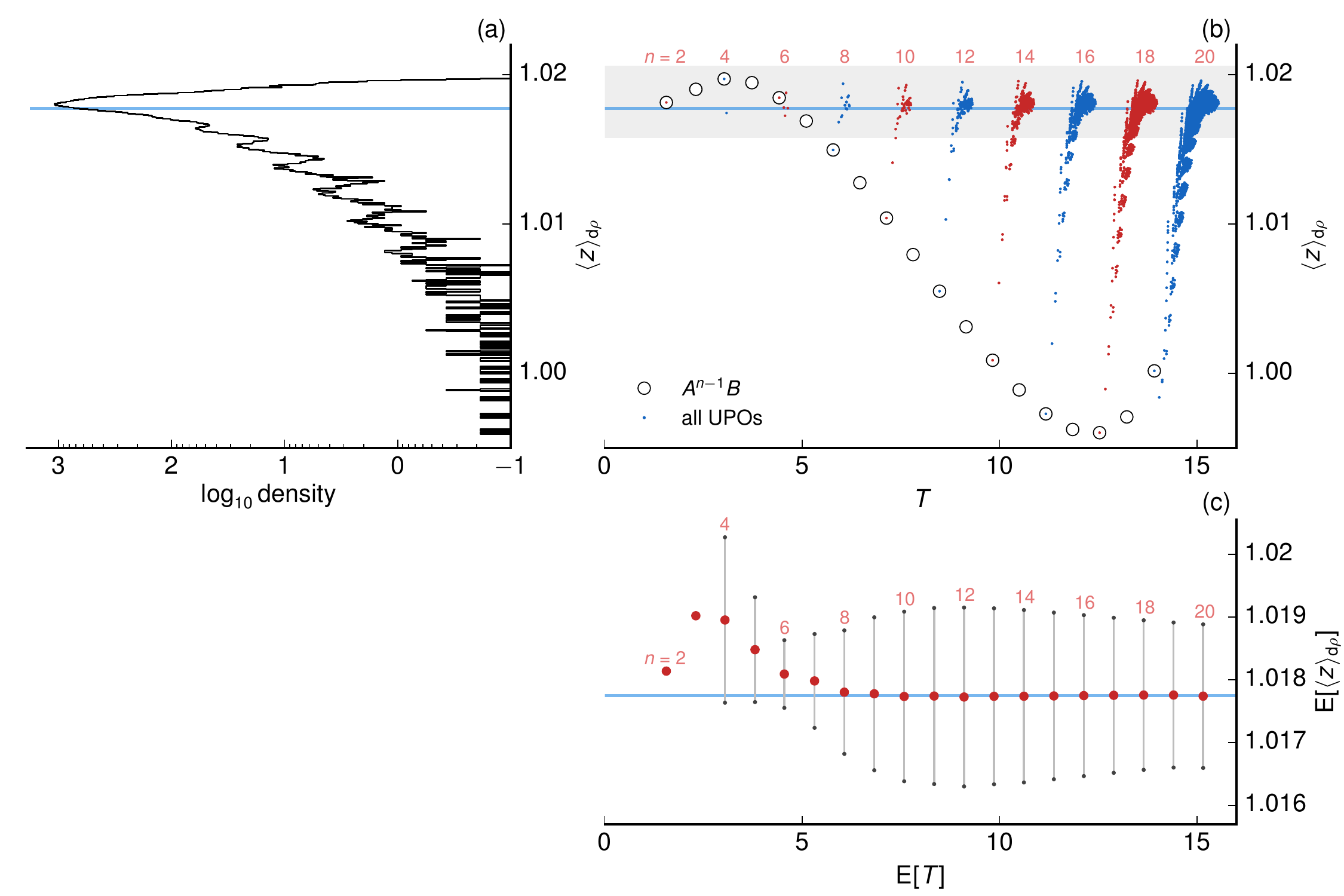}
	\caption{Panel (a): normalised histogram of $\langle z\rangle_{\mathrm{d}\rho}$. The horizontal bar denotes the mean of the distribution. Panel (b): scatter plot of period and gradient of UPOs with even cycle number. Circles identify UPOs with symbol sequence $A^{n-1}B$, for all $n$. The grey band defines the range displayed in panel $(c)$. Panel (c): expectation of gradient and period. The vertical bars span the standard deviation of the distributions for each $n$.}
	\label{fig:lorenz-gradient}
\end{figure}
\subsection{Statistics of UPO gradients}
{The purpose of this section is to analyse the statistical distribution of the sensitivity across UPOs of the Lorenz equations. We will sometimes make use of the expectation $\textsc{E}[\,\cdot\,]$, defined as the equally weighted average over all UPOs. This average has no formal justification, but we use it here and in the following sections primarily to give a quantitative measure of the distribution, for instance, to quantify the standard deviation of the sensitivity across UPOs of same cycle length.} 

At the standard parameters, we find a set of 107406 unique orbits with cycle length $n \le 20$, about 3\% less than the number of orbits (111011) known to compose the symbolic dynamics for $n\le 20$ \cite{Viswanath:2003gb}. This set includes symmetry-invariant UPOs {and pairs of symmetry-equivalent orbits, where one orbit is generated from the other via action of the symmetry.} Uniqueness is verified using the graph-based approach described in \cref{app:graph}.  UPOs are refined iteratively until the estimated error on the orbit baricenter $\epsilon_{\mathbf{h}}$ in subsequent iterations is smaller than $10^{-6}$ (see \cref{app:graph} for details), with a typical time step of $5\cdot10^{-4}$. The gradient $\langle z \rangle_{\mathrm{d}\rho}$ is also typically accurate to this precision.

\cref{fig:lorenz-gradient}-(a) shows the normalised histogram of $\langle z \rangle_{\mathrm{d}\rho}$ constructed by binning the range between 0.995 and 1.02 with 500 bins. No data point falls outside this range. The distribution of the gradients is remarkably narrow; the right tail falls abruptly while the left tail decays more gently, displaying several humps. 
{The gradient expectation (horizontal line in the figure) is \mbox{$\textsc{E}[\langle z\rangle_{\mathrm{d}\rho}] = 1.0177$}, with standard deviation 0.00117.} In \cref{fig:lorenz-gradient}-(b) the scatter of the orbits' period $T$ and gradient $\langle z \rangle_{\mathrm{d}\rho}$ is reported for UPOs of even cycle number $n$. 

The distribution of gradients is structured. Data points are organised in clusters, causing the humps in the left tail of the histogram of \cref{fig:lorenz-gradient}-(a). The subset of UPOs with symbol sequence pattern $A^{n-1}B$, open circles, shown for $2\le n \le20$, displays a wavy pattern and produce extreme values of the gradient, up to $n=19$, after which the orbit $A^{18}B^2$ has the lowest gradient. Tracing UPOs with such pattern for $n > 20$ and examining the asymptotics of their sensitivity would be interesting and is left to future analysis. \cref{tab:lorenz} reports the period $T$, the period average of $z$ and the gradients of these two quantities for such a subset. 

The long orbits can occasionally predict gradients significantly lower than average, causing the gentle decay of the distribution of in \cref{fig:lorenz-gradient}-(a). Nevertheless, the standard deviation of the gradient across UPOs of same cycle length seems to decrease with $n$. This is illustrated in panel (c), reporting the expected gradient $\textsc{E}[\langle z \rangle_{\mathrm{d}\rho}]$ and the expected standard deviation (vertical bars) as a function of the expected period $\textsc{E}[T]$ across orbits of same cycle length $n$. 
\begin{table}[htpb]
	\small
	\caption{Characteristics of the first nineteen UPOs with symbol sequence $A^{n-1}B$. All reported digits are converged. Data points in the third and last columns correspond to the open circles in \cref{fig:lorenz-gradient}-(b).}
	\label{tab:lorenz}
	\centering
	\begin{tabular}{|r|c|c|c|c|c|}
	\hline
	   n & sequence  & $T$          & $T_{\mathrm{d}\rho}$ &  $\langle z \rangle $ & $\langle z \rangle_{\mathrm{d}\rho}$ \\
	\hline
	   2 & $AB$      & \phantom{0}1.5586522 & -0.0392528 & 23.0421039 & 1.0181392\\
	   3 & $A^{ 2}B$ & \phantom{0}2.3059072 & -0.0573500 & 23.2322548 & 1.0190202\\
	   4 & $A^{ 3}B$ & \phantom{0}3.0235837 & -0.0740579 & 23.4646321 & 1.0197131\\
	   5 & $A^{ 4}B$ & \phantom{0}3.7256417 & -0.0899305 & 23.6658224 & 1.0194636\\
	   6 & $A^{ 5}B$ & \phantom{0}4.4177664 & -0.1051958 & 23.8338832 & 1.0184538\\
	   7 & $A^{ 6}B$ & \phantom{0}5.1030384 & -0.1199824 & 23.9742312 & 1.0168975\\
	   8 & $A^{ 7}B$ & \phantom{0}5.7834108 & -0.1343760 & 24.0920415 & 1.0149585\\
	   9 & $A^{ 8}B$ & \phantom{0}6.4602592 & -0.1484423 & 24.1913462 & 1.0127595\\
	  10 & $A^{ 9}B$ & \phantom{0}7.1346327 & -0.1622367 & 24.2751574 & 1.0103966\\
	  11 & $A^{10}B$ & \phantom{0}7.8073881 & -0.1758120 & 24.3456991 & 1.0079511\\
	  12 & $A^{11}B$ & \phantom{0}8.4792710 & -0.1892228 & 24.4045969 & 1.0054978\\
	  13 & $A^{12}B$ & \phantom{0}9.1509718 & -0.2025301 & 24.4530089 & 1.0031120\\
	  14 & $A^{13}B$ & \phantom{0}9.8231719 & -0.2158068 & 24.4917076 & 1.0008768\\
	  15 & $A^{14}B$ &           10.4965892 & -0.2291449 & 24.5211223 & 0.9988917\\
	  16 & $A^{15}B$ &           11.1720295 & -0.2426673 & 24.5413439 & 0.9972860\\
	  17 & $A^{16}B$ &           11.8504579 & -0.2565465 & 24.5520879 & 0.9962404\\
	  18 & $A^{17}B$ &           12.5331044 & -0.2710385 & 24.5525992 & 0.9960262\\
	  19 & $A^{18}B$ &           13.2216410 & -0.2865488 & 24.5414594 & 0.9970803\\
	  20 & $A^{19}B$ &           13.9185025 & -0.3037700 & 24.5162058 & 1.0001659\\
	\hline          
	\end{tabular}
\end{table}

We have used linear regression of long-time averages obtained from integration of the equations for twenty one values of $\rho$ at equally sampled points in the interval across \mbox{$\rho=28\pm 0.25$}, to approximate the gradient $(\overline{z})_{\mathrm{d}\rho}$. Accurate averages of $z$ were obtained by augmenting the equations with the quadrature equation $\dot{\xi}(t) = (z(t) - \xi(t))/t$ (the variable $\xi(\tau)$ is the cumulative average $\xi(\tau) = 1/\tau \int_0^\tau z(t)\, \mathrm{d}t = \overline{z}^{\,\tau}$) and integrating the augmented system using a linear multi-step, variable order integrator with relative and absolute tolerance of $10^{-7}$ from the Sundials suite of integrators \cite{Hindmarsh:2005hg}.


We repeated this approach over 30 trials, using random initial conditions and discarding initial transients. At $\tau=10^6$, the average value of $(\overline{z}^{\,\tau})_{\mathrm{d}\rho}$ across the trials was equal to 1.0027, with a standard deviation of 0.0012. {Different values of the gradient $(\overline{z})_{\mathrm{d}\rho}$ have been reported by other authors, e.g., 0.96 in \cite{LEA:2000cr} and 1.01$\pm$0.04 in \cite{Wang:2013cx}.}
Overall, this approximation lies at the very end of the left tail of \cref{fig:lorenz-gradient}-(a). This is quite far   from where the vast majority of UPOs are located, in particular the short orbits. 

{It is speculated here that} {the mismatch between the numerical approximation of the gradient and the distribution of \cref{fig:lorenz-gradient}-(a)} by the fact that the numerical calculation of statistical quantities over long trajectories in chaotic dynamical systems is a delicate subject \cite{Sauer:1997fp}.   In absence of hyperbolicity, as for the Lorenz equations at standard parameters \cite{Sparrow:2012hi}, ``there is no fundamental reason for computer simulated long-time statistics to be even approximately correct'' \cite{Sauer:2002jc}. In stark contrast, periodic trajectories have desirable convergence properties. Computations in finite precision arithmetic on periodic orbits are \emph{globally} insensitive since the catastrophic temporal propagation of the effects of floating-point errors is constrained by the need to satisfy the periodic boundary conditions of the problem, i.e. robustness is built in the numerical algorithm. Hence, periodic trajectories identified from numerical computation converge when the time step is refined and so do period averages and their gradients, regardless of the orbit length and its stability properties.

\section{Feedback control in Kuramoto-Sivashinsky equation}\label{sec:kuramoto}
In this section, we consider a sensitivity problem for UPOs of the Kuramoto-Sivashinky equation (KSEq). The KSEq is a well known one-dimensional partial differential equation exhibiting complex spatio-temporal behaviour. It arises in a variety of physically relevant context, e.g. reaction diffusion systems \cite{Kuramoto:1978vz} and the dynamics of liquid films falling along a wall \cite{Sivashinsky:1980wg}. A fairly large number of numerical and theoretical investigations on bifurcation, chaos and control \cite{Armaou:2000bb, Gomes:2015gf} have used the KSEq as a model problem. Of particular interest in the current context are early \cite{Christiansen:1997gd} and more recent \cite{Lan:2008kg, Cvitanovic:2010gm} investigations on UPOs of the KSEq.


The equation considered here is 
\begin{equation}\label{eq:ks_equation}
v_{\partial t}(x, t) = (v^2)_{\partial x}(x, t) - v_{\partial^2 x}(x, t) - \kappa v_{{\partial^4 x}}(x, t) + h(x, t)
\end{equation}
where $v(x, t)$ is the scalar field variable, $x$ denotes space, $t$ is time and the subscript $(\cdot)_{\partial^n}$ denotes the $n$-th partial derivative with respect to time or space.
{The additional source term $h(x, t)$ in \cref{eq:ks_equation} represents the distributed volume force actuation that is used to control the system, via a to-be-determined feedback controller $\mathcal{G}$ such that
\begin{equation}\label{eq:controller}
h(x, t) = \mathcal{G}(v(x, t)).
\end{equation}
In such settings, \cref{eq:ks_equation} is still an autonomous system, parametrised by a set of controller variables.}
The equation is parametrised by the diffusivity constant $\kappa$, while the domain length is fixed to $2\pi$. This set up is equivalent to fixing the viscosity, $\kappa=1$, and varying the length of the domain, where the equivalence is $L = 2\pi/\sqrt{\kappa}$, \cite{Lan:2008kg}. 
We consider $2\pi$-periodic, odd solutions, satisfying $v(x+2\pi) = v(x)$ and $v(-x) = -v(x)$. Considering odd solutions reduces the continuous translational symmetry 
\begin{equation}
\tau^\dagger_{\Delta} : v(x, t) \rightarrow v(x+\Delta, t),\;\Delta\in\mathbb{R}
\end{equation}
to a discrete shift-by-$\pi$ symmetry 
\begin{equation}\label{eq:symmetry-pde}
	{S^\dagger : v(x, t) \rightarrow v(x+n\pi, t),\; n\; \mathrm{odd}.}
\end{equation}
The energy density 
\begin{equation}\label{eq:energy-density}
	K^\dagger(t) = \displaystyle K^\dagger(v(x, t)) = \frac{1}{4\pi} \int_0^{2\pi} v(x, t)^2\,\mathrm{d}x
\end{equation}
is the target observable for control. We wish to calculate the gradient $\langle K^\dagger \rangle_{\mathrm{d}\mathcal{G}}$, i.e., the sensitivity of the period averaged cost with respect to the controller $\mathcal{G}$ driving the actuation. For a given UPO, the controller $-\lambda\langle K^\dagger \rangle_{\mathrm{d}\mathcal{G}}$, for a scalar $\lambda > 0$, has the effect of displacing/distorting the UPO in state space in a way that is optimal for the reduction of $\langle K^\dagger \rangle$. No other choice of $\mathcal{G}$ results in a larger variation of the period average, at least in a linear sense. 
The idea is different from previous chaos control studies whereby an UPO with desirable period averaged characteristics is stabilised by feedback \cite{Shinbrot:1993ue}.
Here, UPOs serve as special trajectories where the sensitivity of the nonlinear chaotic state with respect to the feedback can be extracted exactly. When such information is available for many UPOs, insight can be gained into how the attractor and its statistical moments change under control. Empirically, if most UPOs display similar sensitivity, varying the controller in the direction of the expectation might result in an overall favourable displacement of the attractor. The dynamics will still be chaotic but would be mitigated by actuation, resulting in a reduction of the average cost. 

A spectral Fourier discretisation is used, where
\begin{equation}\label{eq:v-fourier}
	v(x, t) = \sum_{k=-N}^N i v_k(t) e^{i k x},
\end{equation}
$i=\sqrt{-1}$, $k$ is the integer wave number and the symmetry $v_{-k} = -v_k$ applies, as Fourier coefficients are imaginary, $v_k \in \mathbb{R}$, $v_0=0$, for odd functions. Galerkin projection results in 
\begin{equation}\label{eq:galerkin-ks}
	\dot{v}_k(t)  = (k^2 - \kappa k^4)v_k(t) - k \sum_{m=-N}^N v_m(t) v_{k-m}(t) - h_k(t),\quad k = 1,\ldots, N,
\end{equation}
where the control variables $h_k(t) \in \mathbb{R}$ arise from a discretisation of $h(x, t)$ analogous to \cref{eq:v-fourier}. Numerically, we solve \cref{eq:galerkin-ks} by evaluating the convolution sum directly and using the  variable order, multi-step stiff integrator from the Sundials suite \cite{Hindmarsh:2005hg} for time marching, typically setting relative error tolerances to $10^{-6}$. 

With $\mathbf{v} = (v_1, v_2, ..., v_N)^\top \in \mathbb{R}^N$ denoting the state variables vector, and similarly for the {control variables} $\mathbf{h}$, the finite-dimensional equivalents of the energy density \cref{eq:energy-density} and the symmetry transformation \cref{eq:symmetry-pde} are  
\begin{equation}
	K(t) = K(\mathbf{v}(t)) = \mathbf{v}^\top(t)\cdot\mathbf{v}(t)
\end{equation}
and
\begin{align}\label{eq:symmetry-ode}
  S :  v_k \rightarrow \left\{
  \begin{aligned}
	 v_k &\mathrm{\;if\;}k\mathrm{\;is\;even}\\
	-v_k &\mathrm{\;if\;}k\mathrm{\;is\;odd}
  \end{aligned}
  \right .
\end{align}
respectively, where  $K$ is invariant under $S$. Assuming linearity of $\mathcal{G}$, spatial discretisation yields
\begin{equation}
	{\mathbf{h}(t) = \mathbf{G}\mathbf{v}(t)}
\end{equation}
where the entries of $\mathbf{G} \in \mathbb{R}^{N \times N}$, the control gains, parametrise the dynamics and the UPOs of \cref{eq:galerkin-ks}. Sensitivity results will be reported in matrix form, with the gradient \mbox{$\langle K \rangle_{\mathrm{d}\mathbf{G}} \in \mathbb{R}^{N \times N}$} and individual entries denoted as $\langle K \rangle_{\mathrm{d}G_{km}}$. However, notation for operations such as inner products, norms, etc. involving the gradient $\langle K \rangle_{\mathrm{d}\mathbf{G}}$ will be used as if it was a row vector. Equation \cref{eq:galerkin-ks} can be rewritten as
\begin{equation}\label{eq:ks-vector}
	\dot{\mathbf{v}}(t) = \mathbf{f}(\mathbf{v}(t); \mathbf{G}),
\end{equation}
and it is thus formally equivalent to \cref{eq:system}. We use notation such as $\mathbf{v}^o(\mathbf{G})$, with $s$   implicit, to denote a periodic solution of (\ref{eq:ks-vector}) for a particular $\mathbf{G}$, or $\mathbf{v}^o(G_{km})$ for a particular parameter. Similarly, $\langle K(\mathbf{v}^o) \rangle$ and $\langle K(\mathbf{G}) \rangle$ are short hands for $\langle K(\mathbf{v}^o(\mathbf{G})) \rangle$ used to denote the period average of an UPO for a particular $\mathbf{G}$, with implicit $\mathbf{G}$ and $\mathbf{v}^o$, respectively, depending on the context.

\begin{figure}[tbp]
	\centering
	\includegraphics[width=1\textwidth]{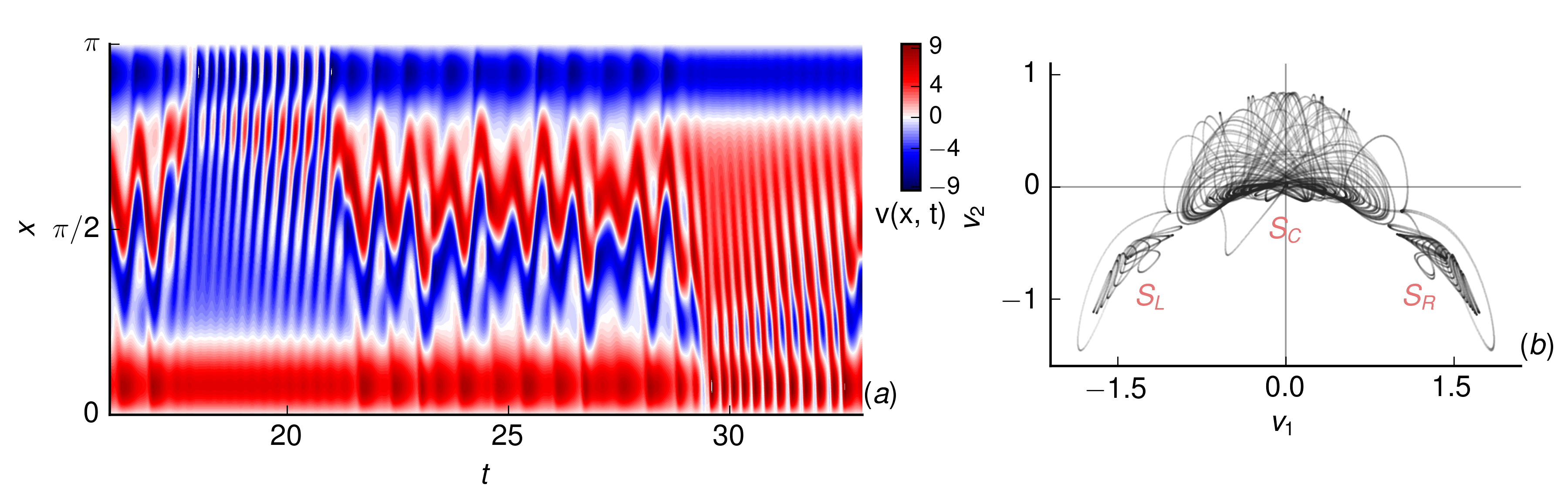}
	\caption{{Panel (a): Spatio-temporal evolution of a chaotic realisation of the uncontrolled KSEq at $\kappa = (2\pi/38.5)^2$. Panel (b): $(v_1, v_2)$ projection of a long trajectory.} }
	\label{fig:chaotic-realisation}
\end{figure}

We investigate control in a ``weakly turbulent'' regime at $\kappa = (2\pi/39)^2$. This value corresponds to a domain size slightly larger than what studied in \cite{Lan:2008kg}, where $L=38.5$. At $L=38.5$, the post-transient dynamics lands onto an attractor composed of a ``center'' region ($S_C$) and two side regions ($S_{L/R}$). A short spatio-temporal sequence of the solution at $L=38.5$ {for the uncontrolled system} is reported in \cref{fig:chaotic-realisation}-(a), while panel (b) shows the $(v_1, v_2)$ projections of a longer trajectory, with labels indicating the three regions of the attractor. In this regime, the dynamics is quite intermittent and trajectories can jump and remain in the side regions for quite some time, before returning to the center region.

In the center region, visited in this realisation during $22\lesssim t \lesssim29$, the solution displays two steady boundary layers, whilst most of the spatio-temporally complex activity, a `left-right' wavy motion, takes place in the domain interior. This activity is the result of multiple concurrent instabilities, yet the dynamics is still relatively low dimensional. Linearisation of \cref{eq:galerkin-ks} around the trivial state $v_k=0, k=1\ldots N$, leads to a diagonal system with entries, the eigenvalues, $\lambda_k = k^2 - \kappa k^4$. At $\kappa = (2\pi/38.5)^2$, there are six unstable modes for $k < \sqrt{1/\kappa} = 38.5/2\pi \approx 6.127\ldots$, with $k=4$ the most unstable. This wave number defines the scale of the coherent structures featuring prominently in the center region. In the side regions, $17\lesssim t \lesssim 22$ and $29\lesssim t \lesssim 32$, the character of the solution changes drastically and describes the oscillatory growth of an instability that saturates back to the center region. At $\kappa = (2\pi/39)^2$, considered in what follows, the side regions of the attractor collapse and long trajectories span the center region only. The evolution is thus considerably simpler and not intermittent, facilitating the numerical search of UPOs and the interpretation of sensitivity and control results.  

Truncation of \cref{eq:v-fourier} at $N=32$ (twice the resolution used in \cite{Lan:2008kg}) results in an overall dynamic range of about six orders of magnitude. Robustness of statistical quantities to perturbations in the truncation is a subtle issue \cite{Cvitanovic:2010gm}. Here, the attractor structure remains intact as the resolution is increased/decreased to $N=64/16$. However, the system appears to be structurally unstable with respect to perturbations in the control gains, as discussed in \cref{sec:control}. 

\subsection{Search results}
{A database of $20073$ unique UPOs has been generated. The large number of orbits is used to explore the statistical distribution of key quantities across UPOs, most importantly the sensitivity with respect to control. Like for the Lorenz system, UPOs are either symmetry-invariant, or come into symmetry-equivalent pairs. Only one of the two orbits is stored in the database.} We count orbits using the cycle length $n\ge1$, the number of `up' intersections with the Poincar\'e section $v_1=0, \dot{v_1}(s) > 0$. For practical purposes, we only search and store UPOs with $n\le 9$.
 \begin{figure}[tbp]
	\centering
	\includegraphics[width=0.85\textwidth]{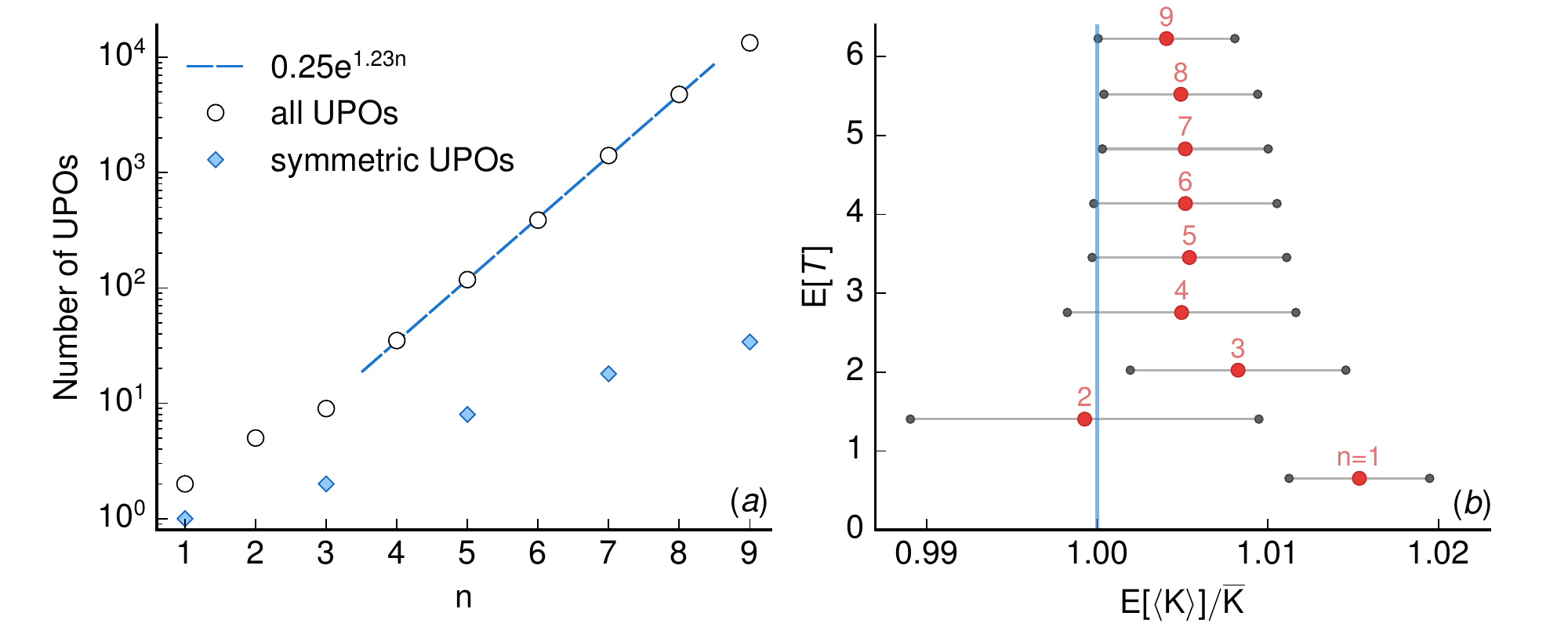}
	\caption{Panel (a): number of orbits as a function of $n$, open circles, and exponential fit for $4\ge n \ge 8$. The blue diamonds denote the number of symmetry invariant UPOs. Panel (b): expectation of the normalised period averaged energy density, as a function of the expectation of the period. The horizontal bars span the standard deviation of the period average, for each $n$.}
	\label{fig:topo-means-paper}
\end{figure}

\Cref{fig:topo-means-paper}-(a) shows the number of UPOs found as a function of $n$ (open circles). An exponential fit for orbits with $4\le n \le 8$ provides an estimate of the topological entropy equal to 1.23 \cite{Auerbach:1987baa}. The fact that the data follows closely the exponential fit, at least up to $n=8$, suggests that the vast majority of UPOs of length in this range has been found.   Symmetry-invariant UPOs (blue diamonds, for odd $n$ only) are the {vast} minority (for $n=9$ we found 34 over 13345). 

In panel (b), the expectation of the period averaged energy density across UPOs of same $n$, normalised by the long-time average, is reported as a function of the period expectation (red circles). The horizontal bars show the span of the standard deviation among UPOs, for each $n$. Even short UPOs give remarkably accurate predictions of the average energy density. Longer orbits yield on average more accurate predictions and the standard deviation amongst orbits of same length decreases with $n$. 

\begin{figure}[htbp]
	\centering
	\includegraphics[width=0.95\textwidth]{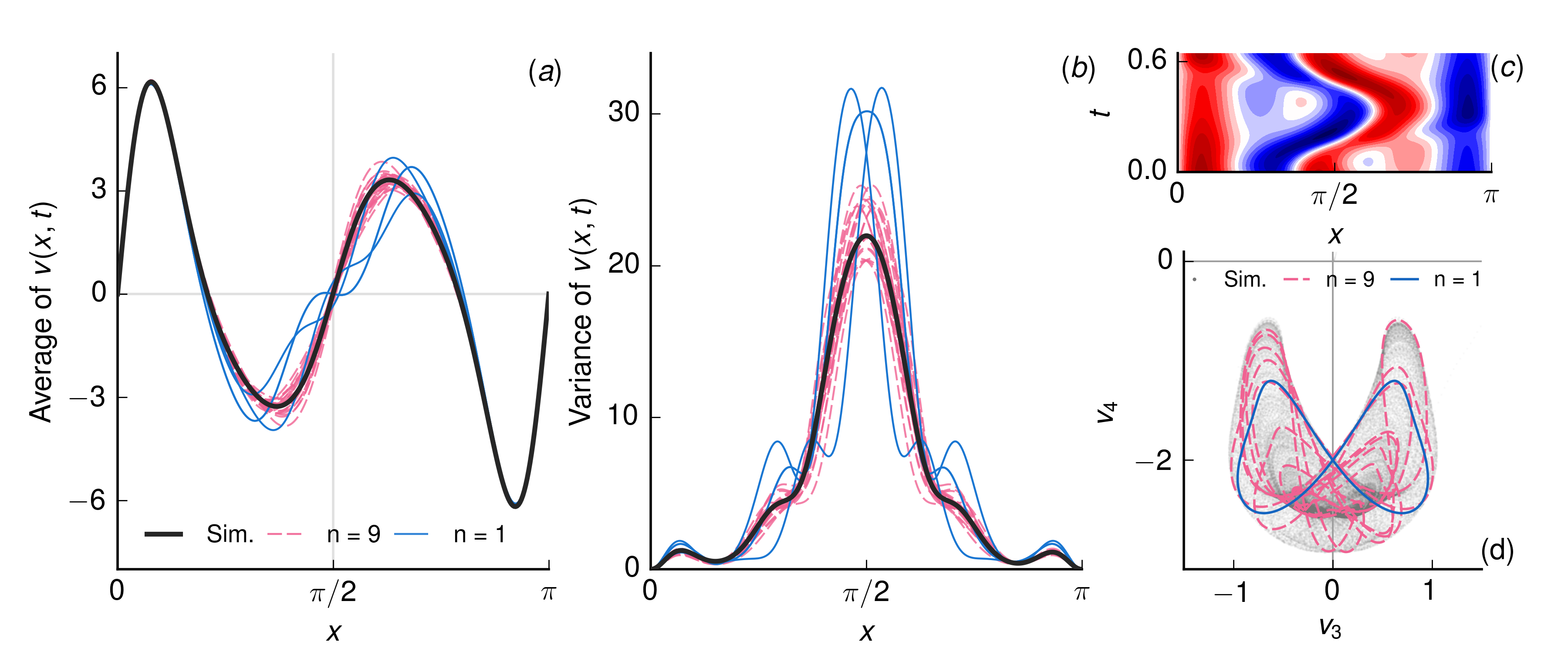}
	\caption{Panel (a): long-time and period averages of $v(x,t)$ for a long chaotic realisation and for selected UPOs of varying length. Panel (b): long-time and period variances. Panel (c): the shortest UPO (same colors as in \cref{fig:chaotic-realisation}). Panel (d): $(v_3, v_4)$ projections of UPOs and long chaotic realisation.}
	\label{fig:mean-and-examples}
\end{figure}


The average spatial structure of {the} chaotic realisation of \cref{fig:chaotic-realisation} is well reproduced by UPOs. This is illustrated in \cref{fig:mean-and-examples}, where statistics for all short ($n=1$, {solid blue lines}) and a subset of ten long ($n=9$, {dashed pink lines}) UPOs are compared to long-time statistics of a chaotic realisation ($T=1000$ thick black line). Panel (a) shows the long-time and period averaged solutions, $\overline{v}$ and $\langle v \rangle$, whilst panel (b) shows the the long-time and period variances $\overline{(v - \overline{v})^2}$ and $\langle (v - \langle v \rangle)^2 \rangle$. The UPOs reproduce particularly well the boundary layers, where the chaotic activity is weaker. These contribute in large part to the overall energy density. The second order statistics in the domain interior, are somewhat less well reproduced, with the $n=1$ UPOs over-predicting the variance at $\pi/2$ by about 40\%. As in \cref{fig:topo-means-paper}, more accurate predictions are obtained for longer UPOs, as these span a larger volume of the attractor and underpin longer segments of chaotic trajectories. This is illustrated in panel (d), where the projections on the subspace ($v_3, v_4$) of a short and a long ($n=1/9$, blue solid/pink dashed line) UPOs are reported, together with a long chaotic trajectory (grey dots) sketching the attractor (this corresponds to the center part $S_C$).

The instantaneous spatio-temporal features of chaotic solutions are also well explained by UPOs. For instance, the shortest UPO ($n$=1), in \cref{fig:mean-and-examples}-(c) describes the ``minimal'' spatio-temporal ``building block'' of the chaotic dynamics and its time scale \cite{Christiansen:1997gd}, the left-to-right motion of the wave in the domain center, while longer UPOs reproduce variations and combinations of this pattern (see e.g. the $n=5$ UPO in \cref{fig:spatial-sensitivity}-(a)).

\begin{figure}[tbp]
	\centering
	\includegraphics[width=1\textwidth]{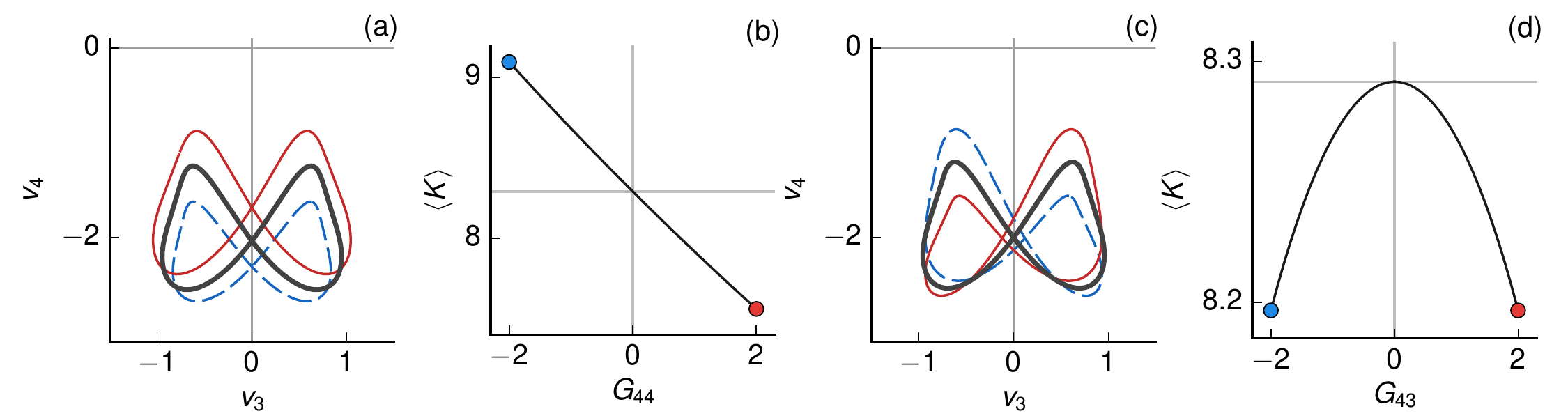}
	\caption{Panels (a) and (c): $(v_3, v_4)$ projections of the original (thick black line) and perturbed UPOs, for perturbations of $G_{44}=\pm 2$ and $G_{43} = \pm 2$, respectively. Negative perturbations correspond to the dashed orbits. Panels (c) and (d): period averaged energy density as a function of the perturbed parameter. Circles denote the value of the parameter at which projections are shown in the corresponding panels. Note that the slope of the bifurcation curves in (b) and (d) at $G_{44}/G_{43}=0$ equal exactly the sensitivity from the adjoint calculation.}
	\label{fig:role-of-symmetries}
\end{figure} 

\subsection{Sensitivity analysis: role of symmetries}
We begin the sensitivity analysis results with a discussion on the role of the symmetry \cref{eq:symmetry-ode} on the sensitivity calculations. System \cref{eq:ks-vector} is {equivariant} under \cref{eq:symmetry-ode}, if the vector field $\mathbf{f}$ and the transformation $S$ commute, i.e. if
\begin{equation}\label{eq:invariance}
	S(\mathbf{f}(\mathbf{v}; \mathbf{G})) = \mathbf{f}(S(\mathbf{v}); \mathbf{G})
\end{equation}
holds. When $\mathbf{G}=0$, \cref{eq:invariance} holds identically. 

Algebraic manipulations omitted here show that \cref{eq:invariance} also holds for perturbations of the control gains $G_{km}$, when $k+m$ is even. When these control gains are varied, symmetry-invariant UPOs move in state space whilst preserving the symmetry. An instance is {shown} in \cref{fig:role-of-symmetries}-(a), for the shortest ($n=1$) UPO. The original UPO (thick black orbit) and two orbits obtained from continuation along $G_{44}$ ($G_{44}=+2/-2$, solid red/dashed blue orbits, respectively) are shown, projected on the $(v_3, v_4)$ subspace. When this parameter is varied, the UPO moves towards/away from the origin for positive/negative perturbations, whilst remaining symmetric. The period averaged energy density varies accordingly, \cref{fig:role-of-symmetries}-(b). In practice, the displacement of UPOs in state space results in the displacement of the entire attractor and thus time averages from long chaotic trajectory vary similarly.

On the other hand, perturbations of control gains $G_{km}$, when $k+m$ is odd, break   \cref{eq:invariance}. Depending on the sign of the perturbation, the symmetry can be broken in one direction or the other. This is illustrated in \cref{fig:role-of-symmetries}-(c), for parameter $G_{43}$. For any perturbation $G_{43}$, the perturbed orbit satisfies globally {$\mathbf{v}^o(s, G_{43}) = S(\mathbf{v}^o(s, -G_{43})),\forall s$}. Because the energy density is invariant under the symmetry, i.e., $K(\mathbf{v}) = K(S(\mathbf{v}))$, it follows that $\langle K(G_{43})\rangle = \langle K(- G_{43}) \rangle$ for all $G_{43}$. Hence, the period averaged energy density is an even function of $G_{43}$ and the gradient $\langle K \rangle_{\mathrm{d}G_{43}}$ must vanish identically at the origin.

\begin{figure}[tbp]
	\centering
	\includegraphics[width=1\textwidth]{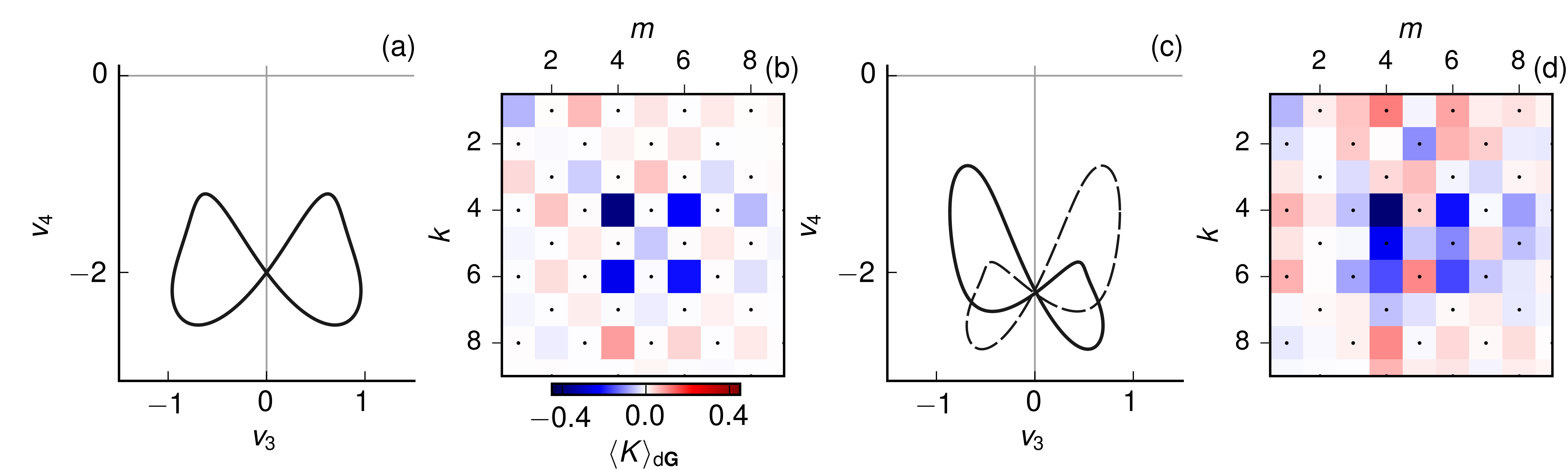}
	\caption{Panels (b)-(d): the gradient $\langle K \rangle_{\mathrm{d}\mathbf{G}}$ for the two shortest UPOs ($n=1$). Panels (a)-(c): their projections on the ($v_3, v_4$) subspace.}
	\label{fig:gradient-projections}
\end{figure}

The effect of symmetries on UPOs that are not symmetry-invariant is illustrated in \cref{fig:gradient-projections}. Panels (a) and (c) show projections on the ($v_3, v_4$) subspace of the three $n=1$ UPOs, the symmetry-invariant UPO, in (a), and the {symmetry-equivalent} pair, in (c). 
The dashed orbit in the latter panel is obtained by application of \cref{eq:symmetry-ode}. Panels (b) and (d) display the gradient $\langle K \rangle_{\mathrm{d}\mathbf{G}}$. The entries of the gradient decay exponentially with $m$ and $k$, hence only the relevant part is shown. As anticipated, for the symmetric orbit the gradient with respect to gains where $k+m$ is odd, indicated by black dots in \cref{fig:gradient-projections}-(b), is identically zero and the gradient display a checker board pattern. 
The orbits in \cref{fig:gradient-projections}-(c) do not have this property and all entries are generally non zero (the figure shows the gradient of the orbit denoted by the solid line). However, the gradient obtained for the symmetric counterpart satisfies 
\begin{eqnarray}
  \left\{
  \begin{aligned}
  	&\langle K(\mathbf{v}^o) \rangle_{\mathrm{d}G_{km}} = &\phantom{+}\langle K(S(\mathbf{v}^o)) \rangle_{\mathrm{d}G_{km}} \quad & \mathrm{for}\; k+m\; \mathrm{even}\\
  	&\langle K(\mathbf{v}^o) \rangle_{\mathrm{d}G_{km}} = &-\langle K(S(\mathbf{v}^o)) \rangle_{\mathrm{d}G_{km}} \quad & \mathrm{for}\; k+m\; \mathrm{odd},
  \end{aligned}
  \right .
\end{eqnarray}
Hence, the joint contribution of the orbits of a {symmetry-equivalent} pair displays the same pattern as symmetry-invariant orbits. 


This result has significant implications for sensitivity analysis and optimisation of chaotic dynamics using, but not limited to, the methods discussed here. The primary reason is that many such systems possess symmetries that would lead to the same behaviour illustrated here. {For instance, many fluid flow configurations of practical interest, e.g. shear flows in channels or pipes, possess continuous symmetries along the homogeneous spatial directions. Hence, a first order sensitivity analysis might not be sufficient to reveal the true influence of system parameters. A second order analysis might be necessary for optimisation over finite amplitude perturbations. The development of second order sensitivity techniques is thus an important avenue for future work.}

\subsection{Sensitivity analysis: statistics of gradients}
\begin{figure}[tbp]
	\centering
	\includegraphics[width=0.72\textwidth]{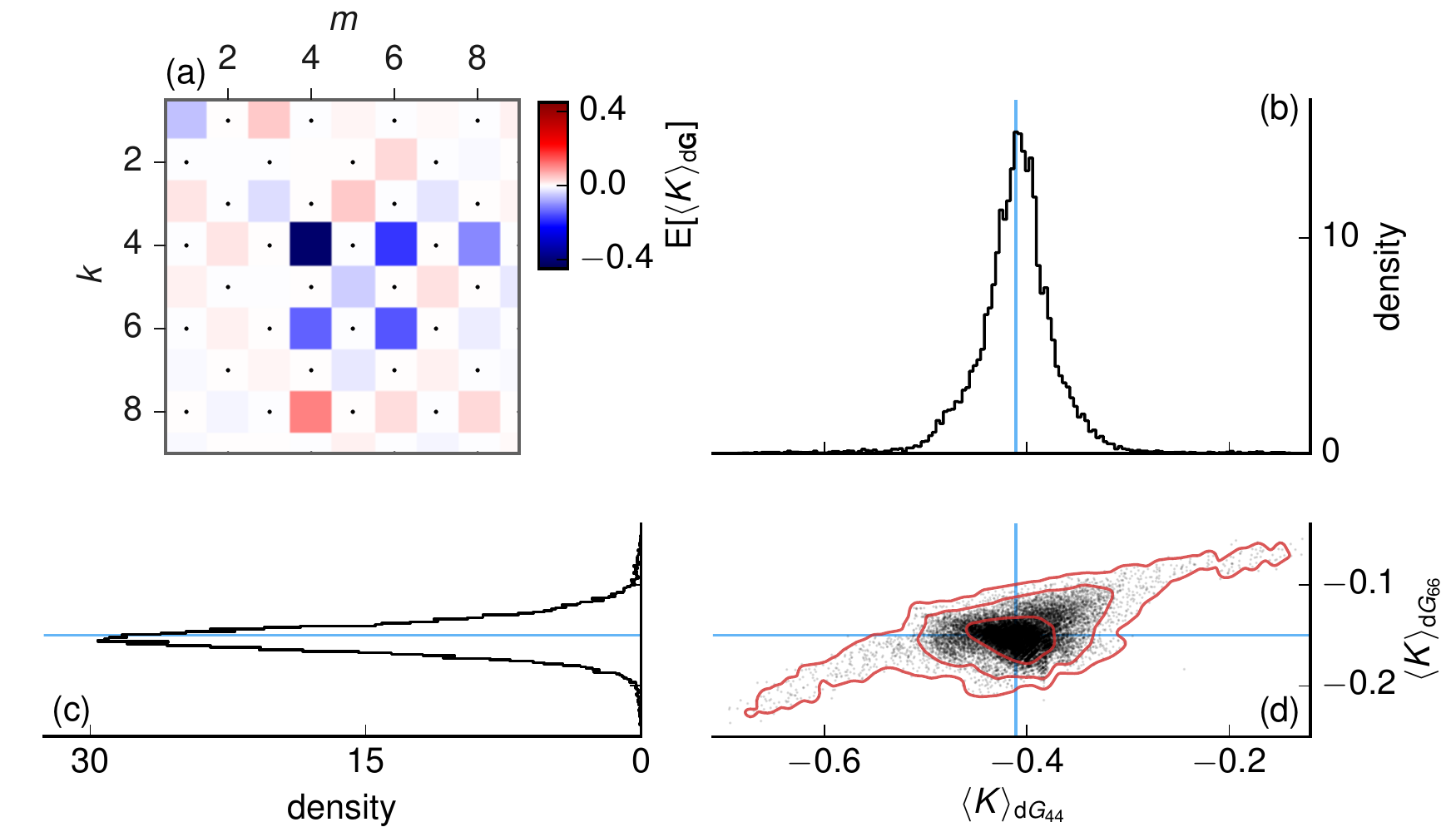}
	\caption{Panel (a): gradient expectation. Panel (d): scatter of components $\langle K \rangle_{\mathrm{d}G_{44}}$ and $\langle K \rangle_{\mathrm{d}G_{66}}$ (black dots). In red, contour lines of the joint probability density function of the two components, for density $1$, $10$, and $100$. Panels (b-c): normalised histograms of $\langle K \rangle_{\mathrm{d}G_{44}}$ and $\langle K \rangle_{\mathrm{d}G_{66}}$, respectively.}
	\label{fig:statistics}
\end{figure}

\cref{fig:statistics}-(a) shows the expectation of the sensitivity $\langle K \rangle_{\mathrm{d}\mathbf{G}}$. The expectation displays the same checker board pattern observed for symmetry-invariant orbits, because opposing contributions from pairs of {symmetric-equivalent} orbits are included in the average. 
 
\begin{figure}[tbp]
	\centering 
	\includegraphics[width=1\textwidth]{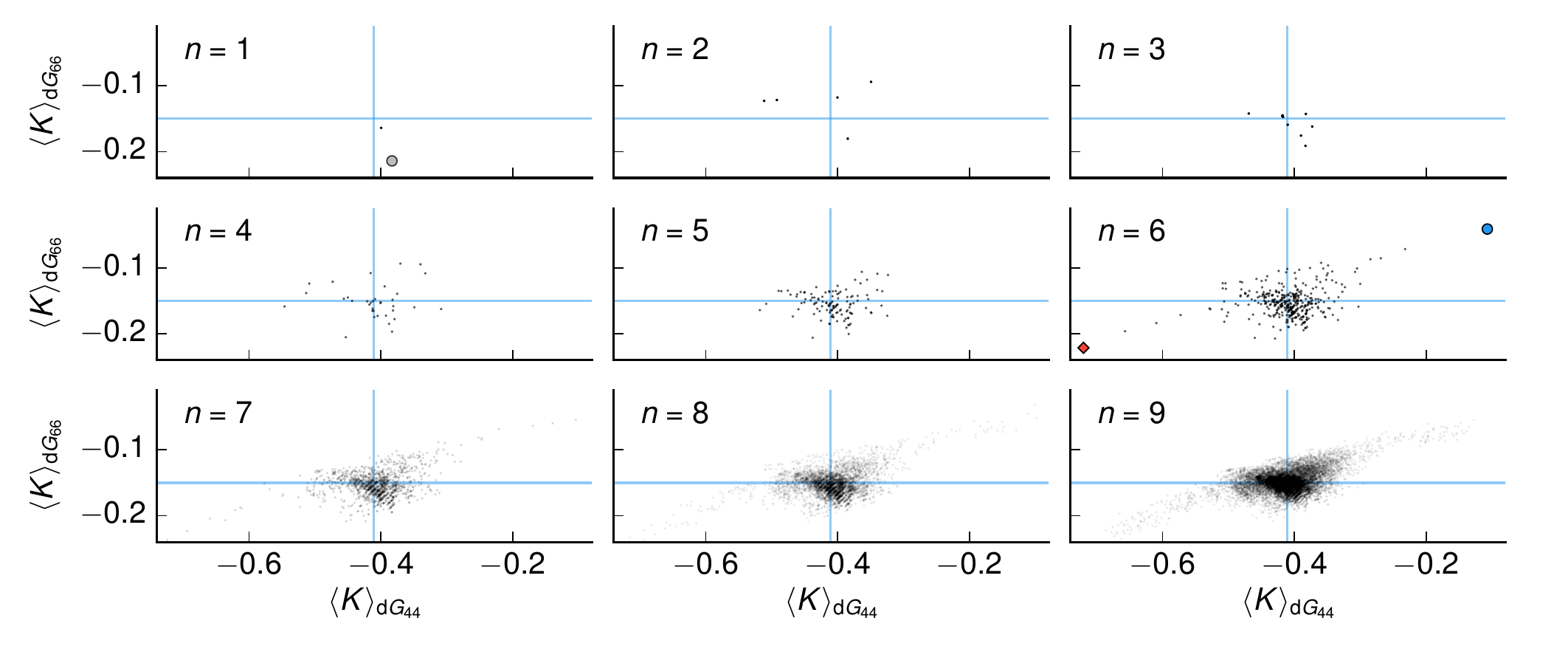}
	\caption{Scatter $\langle K\rangle_{\mathrm{d}G_{44}}$ and $\langle K\rangle_{\mathrm{d}G_{66}}$ across UPOs of different cycle length $n$. The orbits denoted by larger symbols ($n$=1 and 6) are those for which continuation analysis is performed in \cref{fig:gradient-extreme-UPOs}.}
	\label{fig:gradient-with-n}
\end{figure}

In panels (b) and (c), normalised histograms of two components with large magnitude, $\langle K \rangle_{\mathrm{d}G_{44}}$ and $\langle K \rangle_{\mathrm{d}G_{66}}$, are reported. Other components display similar behaviour (distributions for gradients components for odd $k+m$ have zero odd moments). Panel (d) shows the scatter of these two components and three contours of the joint probability density function obtained using a kernel density estimation technique with bandwidth 0.005. Analysis of these and other distributions shows that the vast majority of UPOs predicts approximately the same sensitivity, i.e. gradients are tightly aligned around the expectation of panel (a), denoted by lines in the other panels. 
We attribute this behaviour to the fact that the attractor has a rather simple structure, the system displays a limited range of dynamical behaviour and all UPOs describe variations of the same spatio-temporal pattern. It is argued that in systems with larger number of active degrees of freedom, with higher-dimensional and more convoluted attractors, the scatter of sensitivity among UPOs will be larger. 

\cref{fig:gradient-with-n} shows the scatter of the same two gradient components for orbits of cycle lengths $1\leq n \leq 9$. UPOs with $n \le 5$ appear to be mostly concentrated around the expectation (blue lines), although for $n\ge 6$ a handful of orbits   displays much larger or smaller gradient than average, producing the two long tails of the density distribution of \cref{fig:gradient-projections}-(d).

\begin{figure}[htbp]
	\centering
	\includegraphics[width=1\textwidth]{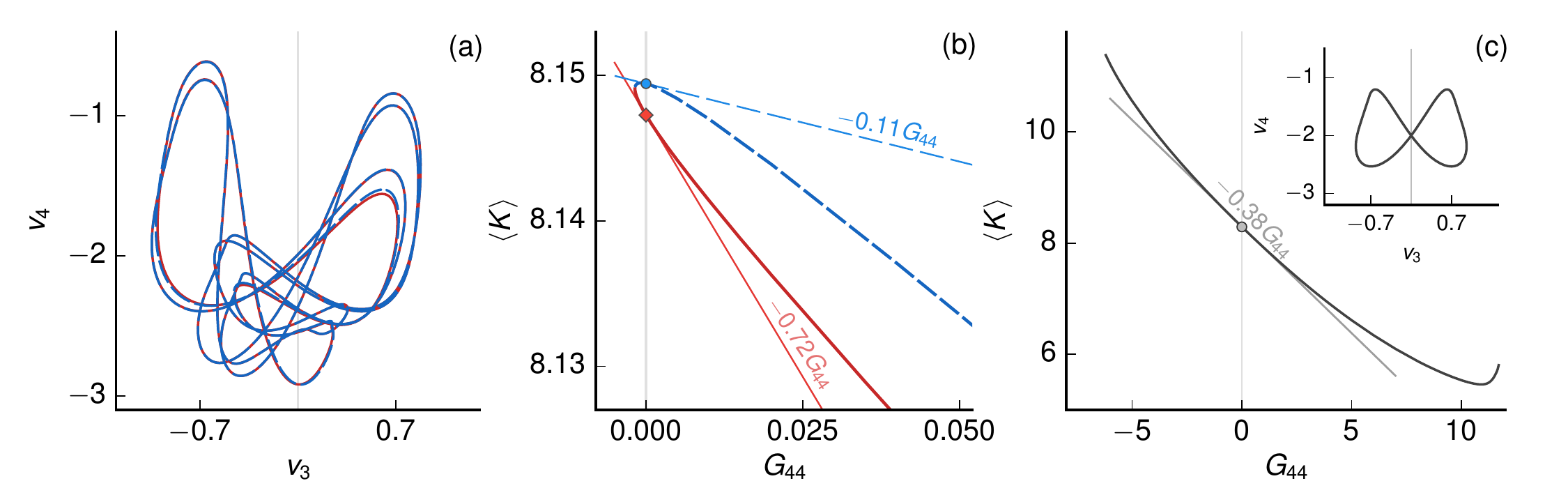}
	\caption{Panel (a): the two $n=6$ UPOs with largest/smallest sensitivity $\langle K \rangle_{\mathrm{d}G_{44}}$. Continuation analysis along $G_{44}$ for these two orbits is shown in panel (b). Here, the two lines tangent to the bifurcation curves have slope equal to what illustrated in the relevant panel in figure (\cref{fig:gradient-with-n}), with numerical value of the gradient denoted by the label. Panel(c): bifurcation curve for the shortest UPO ($n=1$), displayed in the inset in the same panel.}
	\label{fig:gradient-extreme-UPOs}
\end{figure}

We perform continuation along $G_{44}$ for the two $n=6$ UPOs with extreme value of the sensitivity $\langle K \rangle_{\mathrm{d}G_{44}}$ (denoted in the relevant panel of \cref{fig:gradient-with-n} with larger symbols). One would expect that UPOs with such large difference in sensitivity have significantly different characteristics. However, the projections of these two UPOs on the $(v_3, v_4)$ subspace, reported in \cref{fig:gradient-extreme-UPOs}-(a), show that they are actually similar to each other. The continuation analysis, panel (b), reveals that the two orbits actually sit on the upper and lower branches of the same bifurcation curve, originating in a saddle-node bifurcation just before $G_{44} = 0$. The slope of the lines tangent to the bifurcation curve is the sensitivity predicted by the adjoint analysis at $G_{44} = 0$. These two UPOs yield larger/smaller than average sensitivity simply because they sit close to the bifurcation. They survive the continuation process for larger positive values of $G_{44}$ than shown in the figure. Asymptotically, the branches tend to have slopes that agree well with the expectation and with UPOs that are not about to disappear. For instance, for the shortest $n=1$ UPO, panel (c), bifurcation occurs for large absolute values of $G_{44}$, and the gradient is physically relevant. It is argued that other pairs of UPO with larger/smaller than average gradient for $n\ge 6$ display the same behaviour. This has not been checked, as tracing bifurcation curves for all {UPOs} would be too expensive. This is aggravated by the fact that there are many bifurcation parameters in the present problem and saddle-node bifurcations might lurk only in some directions in parameter space, biasing the sensitivity analysis result only in those directions.

These results suggests that care should be taken in interpreting the sensitivity of UPOs. Near a saddle-node bifurcation, the nonlinear BVP \cref{eq:loop-equation} is ill-conditioned: small changes in the parameters result in `large' changes in the solution, distorting the sensitivity results. On the positive side, however, the ill-conditioning is unrelated to the length of the trajectory, e.g. increasing exponentially with time as for the IVP \cref{eq:system}, \cite{Wang:2014hu}. In addition, such bifurcations can be identified with relative ease using standard bifurcation techniques, or with second order sensitivity analysis, as second derivatives would be large close to bifurcations.

\begin{figure}[htbp]
	\centering 
	\includegraphics[width=0.33\textwidth]{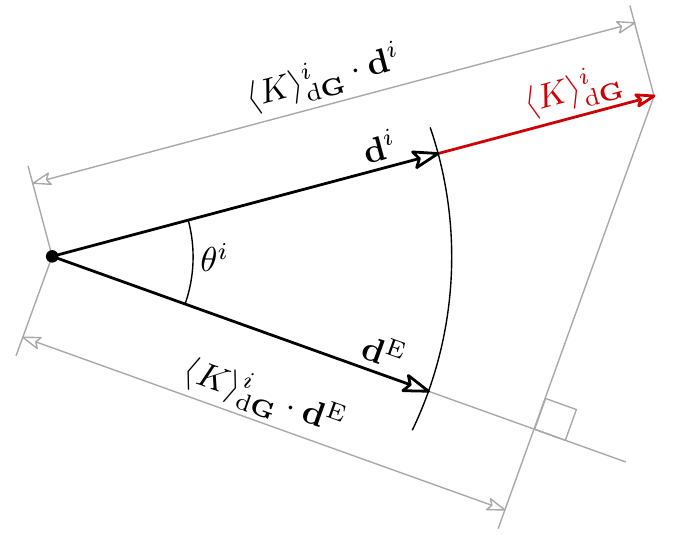}
	\caption{Geometry of gradients and projections.}
	\label{fig:vectors}
\end{figure}

Further gradient statistics relevant for control are now reported. We analyse the geometry and distribution of gradient vectors in parameter space, and quantify their alignment with the gradient expectation {$\mathrm{E}[\langle K \rangle_{\mathrm{d}\mathbf{G}}]$, the equally weighted average of UPO gradients. The aim is to obtain a description of the scatter of UPO gradient vectors in parameter space}. For the $i$-th UPO, the gradient $\langle K \rangle^i_{\mathrm{d}\mathbf{G}}$ is a vector in parameter space pointing in the direction of largest increase of the period averaged cost. Considering \cref{fig:vectors}, let
\begin{equation}
{\mathbf{d}^\mathrm{E} = \mathrm{E}[\langle K \rangle_{\mathrm{d}\mathbf{G}}]^\top / \|\mathrm{E}[\langle K \rangle_{\mathrm{d}\mathbf{G}}]\|\quad \mathrm{and} \quad \mathbf{d}^i = \langle K \rangle_{\mathrm{d}\mathbf{G}}^{i\top} / \| \langle K \rangle_{\mathrm{d}\mathbf{G}}^i\|}
\end{equation}
be the unit vectors pointing along the gradient expectation and the gradient of the $i$-th orbit, respectively. The projections 
\begin{equation}\label{eq:projections}
	\langle K \rangle_{\mathrm{d}\mathbf{G}}^i \cdot \mathbf{d}^\mathrm{E}\quad \mathrm{and} \quad \langle K \rangle_{\mathrm{d}\mathbf{G}}^i \cdot \mathbf{d}^i = \|\langle K \rangle_{\mathrm{d}\mathbf{G}}^i\|
\end{equation}
define the linearised change in the period averaged energy density that would be obtained if control gains were set to $\mathbf{d}^\mathrm{E}$ and $\mathbf{d}^i$, respectively. In other words, these projections define the slope $\langle K\rangle_{\mathrm{d}\lambda}$ of an hypothetical bifurcation curve $\langle K(\lambda) \rangle$ when control parameters are varied as $-\lambda \mathbf{d}^\mathrm{E}$ and $-\lambda \mathbf{d}^i$, respectively, at $\lambda=0$. {The angle $\theta^i = \arccos(\mathbf{d}^i\cdot\mathbf{d}^{\mathrm{E}})$ quantifies the misalignment angle between the gradient of a particular UPO and the expectation.} 
\begin{figure}[htbp]
	\centering
	\includegraphics[width=0.965\textwidth]{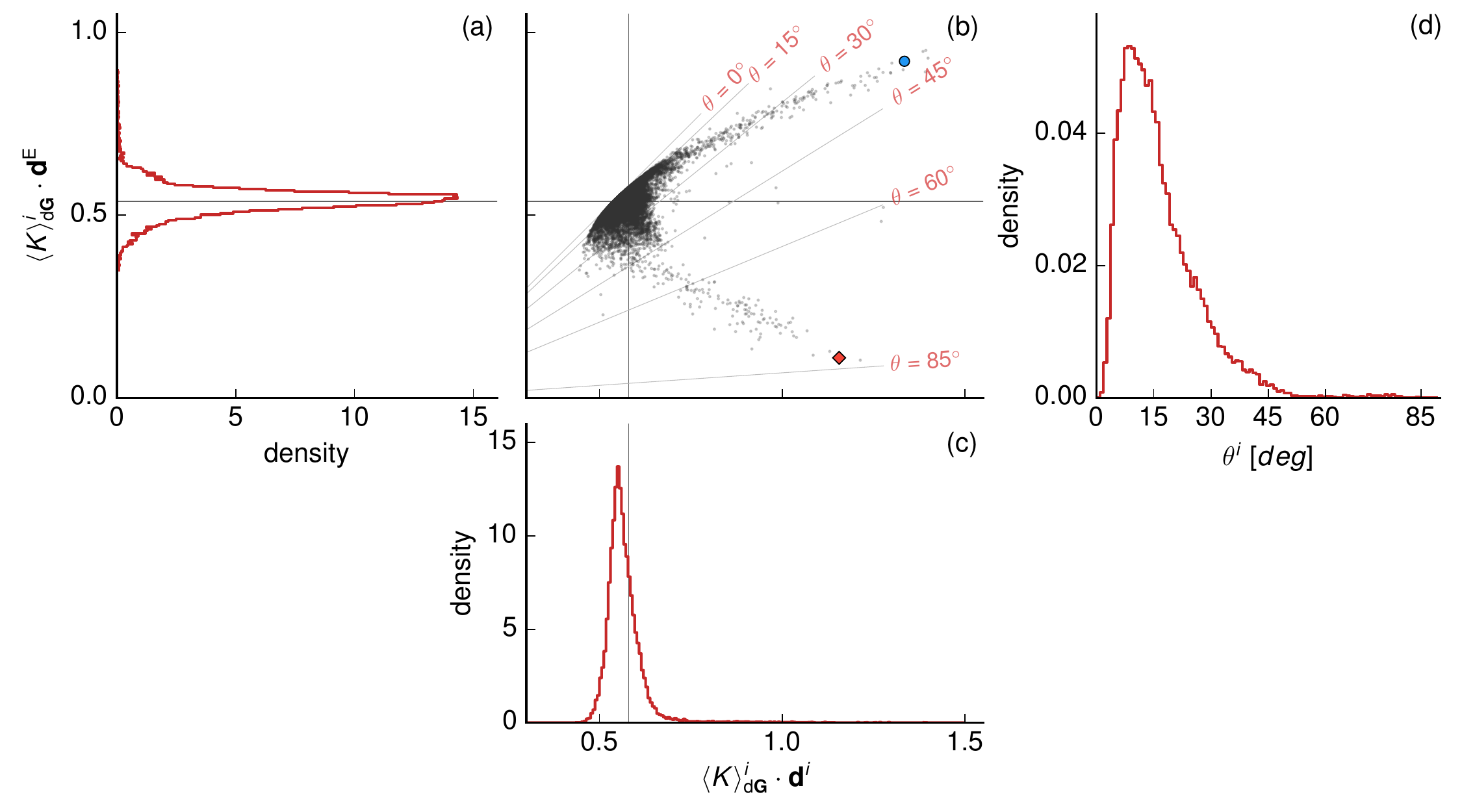}
	\caption{Panels (a), (c) and (d): normalised histograms of the projections \cref{eq:projections} and angle $\theta$, respectively. Panel (b): scatter of the two projections. {Note that the quantity of panel (c) is also the norm of the gradient.}}
	\label{fig:gradient-histograms-multiple-plots}
\end{figure}

Normalised histograms for the two projections \cref{eq:projections} and the angle $\theta$ are reported in panels (a), (c) and (d) of \cref{fig:gradient-histograms-multiple-plots}, respectively. A scatter {plot} of the projections is reported in panel (b). The two marginal histograms in (a-c) are quite narrow, reflecting the compactness of the histograms of individual gradient components in \cref{fig:statistics}. 

These two distributions have weak tails that are better seen in the scatter plot as two tails that extend to large values of the gradient norm $\langle K \rangle_{\mathrm{d}\mathbf{G}}^i \cdot \mathbf{d}^i$. The two closed symbols in (b) correspond to the two $n=6$ orbits described in \cref{fig:gradient-extreme-UPOs}, suggesting that orbits that are close to bifurcation have typically large gradient norm and can point in direction in parameter space quite away from the expected direction, especially orbits on the lower tail of the scatter plot. The distribution of angles in (d) shows that the mode occurs below 10 degrees. More than 90\% of the gradients are within 30 degrees from the expectations and gradients with larger angles correspond to UPOs close to bifurcation. That such a tight distribution is found is remarkable and a clear indication that most UPOs predict a similar sensitivity with respect to actuation. If gradient had random directions, they would be mostly orthogonal in such a high dimensional parameter space ($N^2 = 1024$).
 
\subsection{Feedback control}\label{sec:control}
\begin{figure}[htbp]
	\centering
	\includegraphics[width=0.95\textwidth]{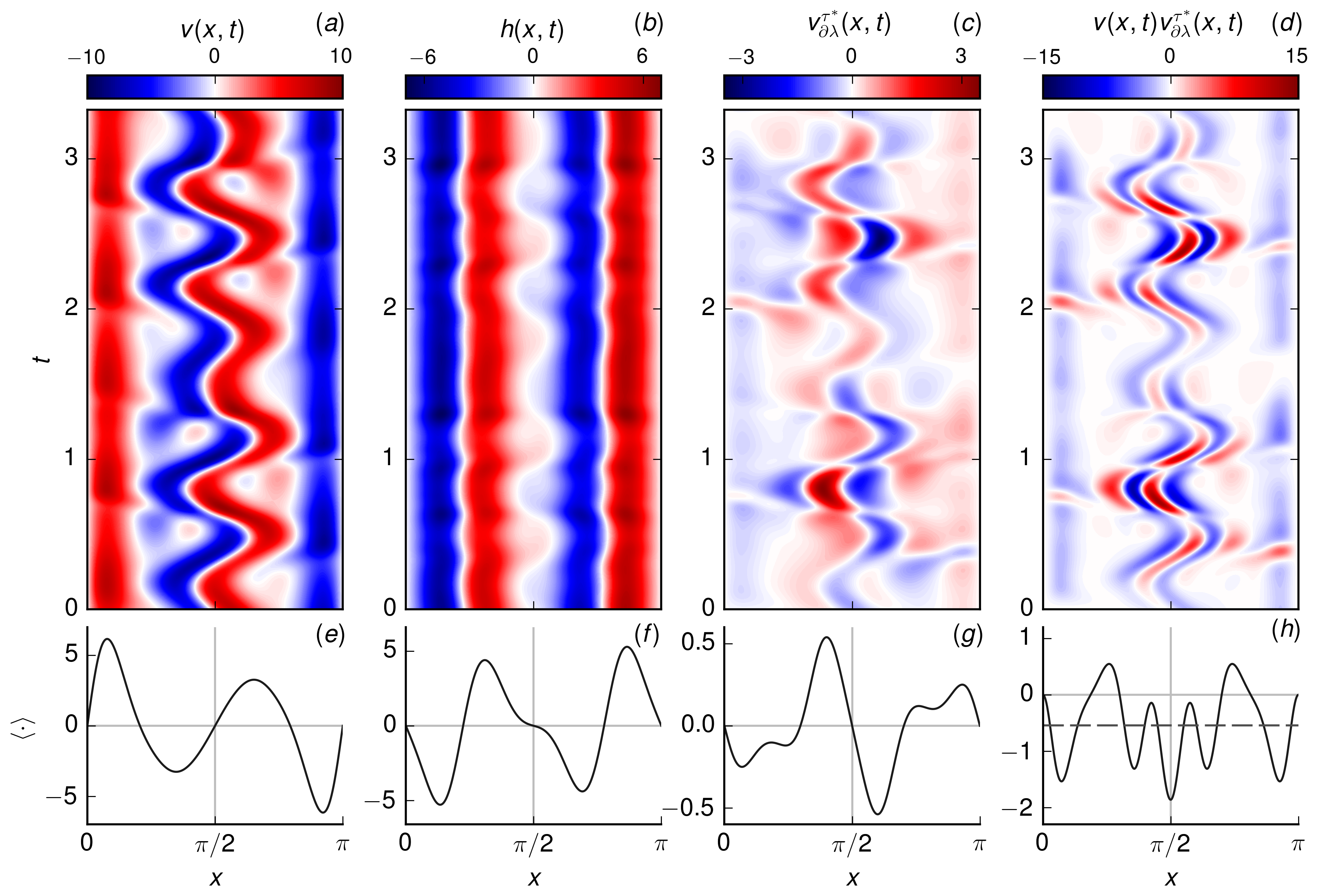}
	\caption{Control mechanisms identified by the sensitivity analysis, illustrated on a symmetric $n=5$ UPO, panel (a). Panel (b): spatio-temporal control distribution. Panel (c) least-squares sensitivity of the solution. Panel (d): linearised change in the energy $v(x, t)^2/2$. The bottom panels show the time-averaged, space-dependent quantity reported in the corresponding upper panel.}
	\label{fig:spatial-sensitivity}
\end{figure}

We now illustrate relevant mechanisms of control identified by the sensitivity analysis and discuss how these affect the dynamics of controlled chaotic realisations. \cref{fig:spatial-sensitivity} present such results illustrated for an example UPO, one of the eight symmetric $n=5$ UPOs, with period $T\approx3.33$. The space-time evolution of this orbit is shown in panel (a).  We solve the adjoint problem for this orbit, obtain the gradient $\langle K \rangle_{\mathrm{d}\mathbf{G}}$ and then construct a feedback controller $-\lambda \langle K \rangle_{\mathrm{d}\mathbf{G}}/\|\langle K \rangle_{\mathrm{d}\mathbf{G}}\|$. 
Panel (b) shows the actuation spatial distribution that would result upon application of the controller with $\lambda=1$ to the solution $v(x, t)$ in (a). 

In the region adjacent to the domain boundaries, actuation is almost stationary and has opposite sign to the solution of panel (a). Actuation in this region reduces the amplitude of the boundary layer and the associated peak of the energy. In the domain center, where most of the chaotic activity occurs, actuation has a lower intensity and counters the left-right wavy motion in this region. Here the solution changes more significantly under actuation. Panel (c) shows the spatio-temporal distribution of the sensitivity
\begin{equation}
	v_{\partial \lambda}^{\tau^*}(x, t) = \sum_{k = -N}^N (v_{\partial \lambda}^{\tau^*})_k(t) e^{i k x}
\end{equation}
i.e. the linearised change in the solution $v(x, t)$ under actuation. The Fourier coefficients of the sensitivity $(v_{\partial \lambda})_k$ are obtained from a finite difference approximation along the descent direction, and the least squares technique discussed in \cref{sec:least-squares} is then applied. The structure of $v_{\partial \lambda}^{\tau^*}(x, t)$ shows that actuation induces significant changes in the dynamics of the wavy motion in the domain center, and less at the domain boundaries, in spite of the strength of the control in the boundary region. 
Note that the UPO period increases under actuation, i.e. control slows down the {chaotic} activity and increases the time scale of the dynamics (this seems to be a general behaviour as $\omega_{\mathrm{d}\mathbf{G}} \cdot \langle K \rangle^\top_{\mathrm{d}\mathbf{G}} > 0$ for the vast majority of UPOs). 

The space-time distribution of
\begin{equation}
	\displaystyle \frac{\partial(v^2(x, t)/2)}{\partial \lambda} = v(x, t) v_{\partial \lambda}^{\tau^*}(x, t),
\end{equation}
is shown in \cref{fig:spatial-sensitivity}-(d). This quantity is the linearised change in the integrand of the cost function \cref{eq:energy-density}. The space and period average of this quantity coincides with the norm of the gradient $\langle K \rangle_{\mathrm{d}\mathbf{G}}$, indicated in panel (h) with a dashed line. One can notice that the cost does not strictly reduce everywhere in time and space, but there are regions/instants where the energy increases under control. The period average of this change, panel (h), shows that the reduction is consistent near the domain boundaries, but also in the domain center, whereas there are two regions, near $x = \pi/2 \pm \pi/4$ where on average the energy increases.

\begin{figure}[htbp]
	\centering
	\includegraphics[width=0.98\textwidth]{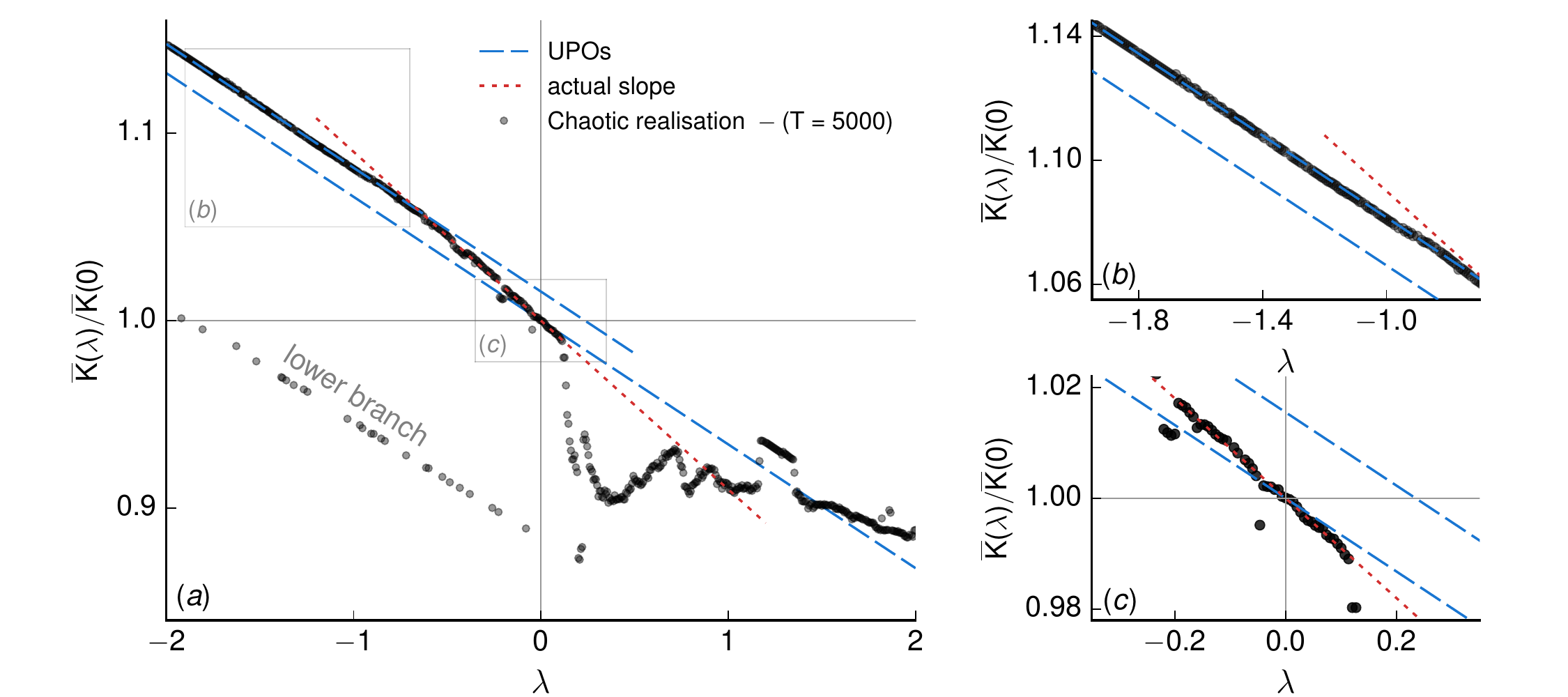}
	\caption{Variation of the normalised long-time average as a function of $\lambda$, the strength of the control, from numerical simulation of the controlled PDE (black dots). The dashed line has slope equal to the norm of the gradient expectation, obtained from UPOs. The dotted line near the origin defines the actual slope of the data points. The two right panels show close-ups of the results in the regions indicated in panel (a).}
	\label{fig:sde}
\end{figure}

We now discuss the effects of actuation on chaotic realisations. We construct the controller $-\lambda \mathbf{d}^\mathrm{E}$, i.e. we vary the control gains in the direction where, on average, the UPOs are most sensitive to. {This is justified empirically by the fact that most UPOs are tightly aligned about this direction. To illustrate the response of the system, we vary $\lambda$ in the range \mbox{$[-2, 2]$}, although it is very likely that many orbits do not survive such a large perturbation.} We then run numerical simulations of the controlled system and monitor the long time ($T=5000$ time units) averaged energy density as a function of the control strength $\lambda$. The results are reported in \cref{fig:sde}, where the vertical axis has been normalised by the long-time average of the uncontrolled system. The dashed lines have slope equal to $\|\mathrm{E}[ \langle K \rangle_{\mathrm{d}\mathbf{G}} ]\|$, the expectation of the reduction of the period averaged cost across UPOs, per unit strength of control $\lambda$ varying the gains in the direction $\mathbf{d}^\mathrm{E}$. This quantity coincides with mean of the distribution of \cref{fig:gradient-histograms-multiple-plots}-(a) and is approximately equal to $0.538$.

The slope associated to the expectation of the UPOs gradients gives a good overall representation of the linear variation of the long-term dynamics of chaotic solutions induced by the feedback control, especially for negative $\lambda$. Closer analysis near $\lambda = 0$, however, shows that the data points are distributed around a straight line (the dotted line) with larger slope, about 0.75, i.e. the local response $\raisebox{1.5pt}{(}\overline{K}\raisebox{1.5pt}{)}_{\partial \lambda}$ (from a data fit) is about 40\% larger than that predicted on average by the UPOs. That the match is not exact is to be expected because the equally weighted average in the expectation has no theoretical basis, but it would be justified, empirically, by the fact that gradients are tightly aligned in parameter space around the expectation. Still the variation of the long-time average with $\lambda$ is much greater than what predicted by the vast majority of UPOs, {much more than for the Lorenz system}. {In particular, the value 0.75 sits on the right tail of the distribution of \cref{fig:gradient-histograms-multiple-plots}-(a). Only a few UPOs reach such large value.}

{Our speculation on this difference is that} the system undergoes significant structural changes around $\lambda=0$. These become particularly evident around $\lambda=0.1$. For $\lambda < -0.5$ the dynamics do not vary significantly and the response of the UPOs appear to give a very good overall prediction of the response of the controlled chaotic dynamics. For $\lambda < 0$, the average can lie on two separate ``branches''. On the lower ``branch'', the dynamics can land and remain, depending on the particular value of $\lambda$ and the initial condition, on the side regions $S_{L/R}$ of an attractor with similar structure to that of $\kappa = (2\pi/38.5)^2$ in \cref{fig:chaotic-realisation}-(b), arising at $\kappa = (2\pi/39)^2$ as a result of control. Long simulations (thousand of time units) indicate that these are disconnected from the ``center'' part $S_C$ (associated to the upper ``branch''), which justifies the fact that the two ``branches'' in \cref{fig:sde} are distinct. The points on the lower branch are the vast minority, though.

For $\lambda > 0$, with the exception of a narrow range around $\lambda \approx 1.2$, the dynamics become highly intermittent. The probability of trajectories landing on the ``side parts'' of the attractor along the duration of the integration increases, and the distribution of the data points for the average bends towards to lower ``branch''. The much larger slope of the data points for the average in the region $0.1 < \lambda < 0.3$ is thus mostly attributed to the increasing intermittent character of the dynamics under stronger actuation and the increasing preference for the system to land of the ``side'' parts, where the energy is lower, rather than a genuinely larger sensitivity of the dynamics on the ``center'' part of the attractor with respect to actuation.

\section{Conclusions}
In this paper we have discussed a robust adjoint sensitivity method for chaotic dynamical systems. The method is based on the idea of formulating the adjoint problem on   time periodic trajectories, rather than on open trajectories. Both types of trajectories are unstable, but in the periodic case the adjoint problem has periodic boundary conditions in time, i.e. it becomes a two-point boundary value problem. The adjoint solution remains bounded at all times and does not exhibit the typical unbounded exponential growth observed in traditional adjoint methods on open unstable trajectories. This special structure enables the exact sensitivity of period averaged statistics with respect to problem parameters to be obtained, \emph{regardless of the length and stability characteristics of the orbit}. Sensitivity information from these trajectories can then reveal important physical features of the parametrised system that would be otherwise hardly accessible.

The solution of the adjoint problem on a periodic orbit requires a strategy different to that used in traditional adjoint analysis for unsteady problems. The backward in time integration is not required and the governing and adjoint equations can in principle be marched forward simultaneously in time. However, methods that have been traditionally used to solve two-point boundary value problems are needed to enforce the time periodic boundary conditions. On the practical side, however, algorithms and software used for the search of periodic orbits can be largely reused.

We have demonstrated the method using two well known low-dimensional chaotic systems: the Lorenz equations at the standard parameters and the Kuramoto-Sivashinsky equation, in a weakly chaotic regime. For the latter problem we have discussed a case that is relevant to applications, e.g. the sensitivity of averages with respect to feedback control. For both problems, we have provided a statistical description of the sensitivity over a large number of UPOs, i.e. over a database of $\sim 10^5$ and $\sim 4\times 10^4$ UPOs, respectively. In both cases, we have found that the scatter of sensitivity across UPOs is rather limited, i.e. most orbits predicted similar response to parameter perturbations. This might be attributed to the relatively low-dimensional nature of these systems. We hypothesise that an increasingly larger scatter will be observed in systems with increasingly larger number of active degrees of freedom, e.g. turbulent fluid systems {at large} Reynolds number. 

We have attempted to estimate the gradient using linear regression of long time averages from chaotic realisations of the perturbed equations and compared the value with the distribution of period averaged gradients from UPOs. The agreement is only fair, not in absolute terms (2\% error on the Lorenz problem), but relatively. The estimated gradient falls far from where the vast majority of UPOs (in particular the short cycles) points to, although it still falls within the tails of the distribution. This might be {speculatively} justified by subtle nonlinear dynamics issues, e.g. intermittency, and to the objective difficulty in obtaining properly converged statistics from long simulations. Further analysis is required. In any case, {the scatter of UPO gradients is lower than that resulting from ensemble average techniques \cite{Eyink:2004gk} or the Least Squares Shadowing method \cite{Wang:2013cx, Wang:2014hu}}.

There are some potential caveats and open questions that need to be highlighted. Firstly, one needs to find a periodic trajectory before its sensitivity can be computed. This has been a relatively straightforward step for the two problems discussed here but it can be an objective challenge for high-dimensional systems. Our plan is utilise these techniques to tackle outstanding control/design problems for three-dimensional Navier-Stokes turbulence, in light of the recent advances in understanding the role and dynamics of exact {coherent structures \cite{Nagata:1990vk,Kawahara:2001ft,GIBSON:2009kp,Kawahara:2012iu,Chandler:2013fi,Budanur:2017jc}}. Although many such structures have now {been} found, {the scalability of the approach to real-world turbulent flows remains an open question.} {In practice, the search of periodic trajectories become increasingly more challenging as the Reynolds number and domain size are increased \cite{Chandler:2013fi}. Sensitivity techniques that do not require temporal periodicity, such as Least-Squares Shadowing \cite{Wang:2014hu}, might be better suited in such a scenario.} {In addition, a sensitivity method for systems with continuous symmetries, typical of shear flows, remains to be developed.}

Secondly, a first order sensitivity analysis might not be sufficient, or not informative enough, for problems with symmetries. In such cases, the sensitivity of period averages for symmetry-invariant orbits, or the joint contribution of groups of orbits where the group is invariant, is identically zero. From an optimisation view point, the averages exhibits saddles along these directions in parameter space and thus second order information becomes necessary for effective optimisation over finite parameter perturbations.

A third caveat is related to the fact that periodic orbits (like any other trajectory) can undergo bifurcations in parameter space. The main issue is not the obvious one regarding the differentiability of period averages \emph{at} a bifurcation point, but rather the fact that the sensitivity for an orbit that is \emph{close} to bifurcating might be much larger/smaller than expected, or physically sound. This might lead to erroneous conclusions about the behaviour of the overall system. For instance, an orbit that is about to disappear/appear in, e.g. a saddle-node-type bifurcation, as a certain parameter is varied will yield a sensitivity with respect to that parameter tending to positive or negative infinity. Clearly, the availability of a large number of periodic orbits, a bifurcation study or a second order sensitivity analysis might alleviate this issue.

Fourthly, there are subtle ``pathological'' features of non-hyperbolic systems, the vast majority of systems in the applied sciences, that suggest care in interpreting the sensitivity results. The primary feature is that statistics might not be smooth functions of the parameters, hence not differentiable with respect to parameters, due to the lack of structural stability \cite{Guckenheimer:up, Ershov:1993el}. What is then the interpretation of the sensitivity of an UPO, if the attractor structure in which this orbit resides can suddenly collapse and statistics change abruptly for infinitesimal parameter perturbations? Another issue is that related to intermittency, as illustrated here for the Kuramoto-Sivashinsky equation. If the degree of intermittency varies as parameters are varied, the statistics of the system might change in a way unrelated to what predicted by UPOs.

We wish to address these issues in future work.

\section*{Acknowledgements}
The author wishes to thank Dr Ati Sharma, Mr Arslan Ahmed and Dr Yongyun Hwang for the fruitful discussions. The Institute of Mathematics and its Applications (ima.org.uk) is also gratefully acknowledged for supporting attendance to the Tenth International Symposium on Turbulence and Shear Flow Phenomena (July 6-9${}^\mathrm{th}$, 2017, Chicago, IL, USA) where part of the present results have been presented.

\appendix
\section{UPO search algorithm}\label{app:upo-search-algo}
We used a Newton-Raphson-type technique traditionally known as \emph{quasi-linearisation} in the literature \cite{Ascher:1994ty} to solve the nonlinear periodic boundary value problem (BVP), equation \cref{eq:loop-equation}, and find UPOs numerically. 
The method is analogous to that reported elsewhere, e.g. \cite{Viswanath:2001ko, Lan:2004ch}. It is reported here primarily to highlight the similarities between the linear time periodic BVP arising in the Newton-Raphson iterations and the linear time periodic BVPs for tangent and adjoint sensitivity \cref{eq:tangent-sensitivity-upo}, \cref{eq:adjoint-problem} and \cref{eq:homo-adjoint-problem}. This similarity enables most of the numerical techniques (and software) to be reused for both search and sensitivity analysis. The numerical approach used to solve these linear time periodic BVPs is described in \cref{app:appendix-numerical-solution}.  





The method starts with a trial solution $\{\mathbf{x}^o_k(s), \omega_k\}, \;k=0$, consisting of a frequency guess $\omega_k$ and a state space loop $\mathbf{x}^o_k(s)$, a smooth candidate periodic trajectory that does not satisfy the governing equations, that is, where the residual 
\begin{equation}
\mathbf{r}_k(s) = \omega_k \dot{\mathbf{x}}^o_k(s) - \mathbf{f}(\mathbf{x}^o_k(s))
\end{equation}
is generally non zero, although $\mathbf{x}^o_k(0) = \mathbf{x}^o_k(2\pi)$.  At each iteration $k = 0, 1, \ldots$ the trial solution is adjusted by calculating a correction $\{\mathbf{y}_k(s), \eta_k\}$ such that the new trial solution $\{\mathbf{x}^o_{k+1}(s), \omega_{k+1}\} = \beta_k \{\mathbf{x}^o_k(s) + \mathbf{y}_k(s), \omega_k + \eta_k\}$ solves the nonlinear periodic BVP exactly, i.e.
\begin{equation}
	\mathbf{r}_{k+1}(s) = (\omega_k + \eta_k)(\dot{\mathbf{x}}^o_k(s) + \dot{\mathbf{y}}_k(s)) - \mathbf{f}(\mathbf{x}^o_k(s) + \mathbf{y}_k(s)) = 0
\end{equation}
where $0 < \beta_k \le 1$ is a damping factor discussed below. Assuming that the correction is small in some sense, linearising the vector field around $\mathbf{x}^o_k(s)$, and neglecting second order terms leads to the linear time periodic BVP
\begin{subequations}\label{eq:linear-bvp}
\begin{empheq}[left={\empheqlbrace}]{align}
	\omega_k \dot{\mathbf{y}}_k(s) =&\, \mathbf{f}_{\partial\mathbf{x}}(s)\cdot\mathbf{y}_k(s) - \mathbf{r}_k(s) - \eta_k\dot{\mathbf{x}}^o_k(s)\\
	\mathbf{y}_k(0) =&\, \mathbf{y}_k(2\pi)\\
	0=&\,\mathbf{y}_k(0)^\top \cdot \mathbf{f}(\mathbf{x}^o_k(0))
\end{empheq}
\end{subequations}
where $\mathbf{f}_{\partial\mathbf{x}}(s) = \mathbf{f}_{\partial\mathbf{x}}(\mathbf{x}_k(s))$ is the Jacobian operator evaluated on the current trial solution and the quantities $\{\mathbf{y}_k(s), \eta_k\}$ are the unknowns. The last row defines a phase locking condition to break the translational invariance along the orbit and ensures that the temporal discretisation of \cref{eq:linear-bvp} does not result in a singular problem. The iterations converge quadratically, but convergence is only local. In this work we used a backtracking technique to improve robustness with respect to poor initial guesses, obtained from scanning long simulation for near recurrences. This consisted in recursively halving the damping factor $\beta_k$ at each iteration from $\beta_k=1$ until $\big\langle | \mathbf{r}_{k+1}(s)|^2 \big\rangle < \big\langle | \mathbf{r}_k(s)|^2 \big\rangle$. Typically, we observed slow convergence in the initial iterations, $\beta_k \ll 1$, and then fast quadratic convergence near the solution. We defined convergence when $\big\langle | \mathbf{y}_k(s)|^2 \big\rangle < 10^{-\alpha}$ and $\big\langle | \mathbf{r}_k(s)|^2 \big\rangle < 10^{-\alpha}$ and $\big | \eta^k \big|  < 10^{-\alpha}$, with $\alpha=13$, typically. Stagnation of the average residual norm and decreasing small damping factors $\beta_k$ were used to detect and stop prematurely iterations from initial guesses that would not converge.


\section{Numerical solution of linear time periodic BVPs}\label{app:appendix-numerical-solution}
The Newton-Raphson iteration problem \cref{eq:linear-bvp}, the tangent sensitivity problem \cref{eq:tangent-sensitivity-upo}, and the adjoint problems \cref{eq:adjoint-problem} and \cref{eq:homo-adjoint-problem} can be all recast into the same standard form for linear time periodic BVPs
\begin{subequations}\label{eq:standard-form}
\begin{empheq}[left={\empheqlbrace}]{align}
	\omega \dot{\mathbf{z}}(s) =&\, \mathbf{A}(s)\cdot\mathbf{z}(s) + \mathbf{b}(s) + \kappa\mathbf{c}(s), \quad s\in [0, 2\pi]\\
	\mathbf{z}(0) =&\, \mathbf{z}(2\pi)\label{eq:standard-form-bc}\\
	\alpha =&\,\mathbf{z}(0)^\top \cdot \mathbf{g}\label{eq:standard-form-plc}
\end{empheq}
\end{subequations}
where $\mathbf{z},\mathbf{b},\mathbf{c} \mathrm{~and~} \mathbf{g}\in\mathbb{R}^{N}$, $\mathbf{A}\in\mathbb{R}^{N\times N}$ and $\omega, \kappa, \alpha\in\mathbb{R}$, and where the unknowns are the solution $\mathbf{z}(s)$ and the parameter $\kappa$. We use a finite difference temporal discretisation of \cref{eq:standard-form} over the $M$-point uniform mesh $s_i = i \Delta s, i = 0, \ldots M-1$, with $\Delta s = 2\pi /M$, and seek the solution  $\hat{\mathbf{z}} = (\mathbf{z}(s_0), \mathbf{z}(s_1), \ldots, \mathbf{z}(s_{M-1}))^\top \in \mathbb{R}^{NM} $ at these points. The fourth-order accurate centered stencil
\begin{equation}
	\dot{\mathbf{z}}(s_i) \approx \frac{1}{12\Delta s}\Big(\mathbf{z}(s_{i-2}) - 8 \mathbf{z}(s_{i-1}) + 8 \mathbf{z}(s_{i+1}) - \mathbf{z}(s_{i+2})\Big) ,\quad i=0, \ldots, M-1.
\end{equation}
is used to approximate the loop derivatives, using the periodicity $\mathbf{z}(s_{i+M}) = \mathbf{z}(s_i)$ as appropriate. This is analogous to the approach used in \cite{Lan:2004ch}. Discretization leads to the set of linear algebraic equations
\begin{equation}\label{eq:discretisation-fourth-order}
	\chi(\mathbf{z}(s_{i-2}) - 8 \mathbf{z}(s_{i-1}) + 8 \mathbf{z}(s_{i+1}) - \mathbf{z}(s_{i+2})) - \mathbf{A}(s_i)\cdot \mathbf{z}(s_i) -  \kappa \mathbf{c}(s_i) = \mathbf{b}(s_i)
\end{equation}
for $i=0, \ldots, M-1$ and with $\chi = \omega/(12\Delta s)$. Addition of the phase locking constraint \cref{eq:standard-form-plc} results in a bordered square system of $NM+1$ linear equations
\begin{equation}\label{eq:bordered-system}
	\begin{pmatrix}
	\mathbf{D} & \hat{\mathbf{c}}\\
	\hat{\mathbf{g}}^\top & 0
	\end{pmatrix} 
	\begin{pmatrix}
	\hat{\mathbf{z}}\\
	\kappa
	\end{pmatrix} =
	\begin{pmatrix}
	\hat{\mathbf{b}}\\
	\alpha
	\end{pmatrix}
\end{equation}
In \cref{eq:bordered-system}, $\hat{\mathbf{g}} = (\mathbf{g}, \mathbf{0}, \ldots, \mathbf{0})^\top \in \mathbb{R}^{NM}$ and the matrix $\mathbf{D} \in \mathbb{R}^{NM\times NM}$, defined as
\begin{equation}
\mathbf{D} = 
\begin{bmatrix*}[r]
	-\mathbf{A}_0        &  8 \chi \mathbf{I}_N & -  \chi \mathbf{I}_N &                       &                      &  \chi \mathbf{I}_N\phantom{_{1-2}}   & -8 \chi \mathbf{I}_N\phantom{_{1-2}} \\ 
	-8 \chi \mathbf{I}_N & -\mathbf{A}_1        &  8 \chi \mathbf{I}_N &  -  \chi \mathbf{I}_N &                      &                                      &    \chi \mathbf{I}_N\phantom{_{1-2}} \\ 
	   \chi \mathbf{I}_N & -8 \chi \mathbf{I}_N & -\mathbf{A}_2        &   8 \chi \mathbf{I}_N & -  \chi \mathbf{I}_N &                                      &                                      \\
	                     &                      &                      &           \ddots      &                      &                     &                                      \\
	                     &                      &  \chi \mathbf{I}_N   &  -8 \chi \mathbf{I}_N & -\mathbf{A}_{3}      & 8 \chi \mathbf{I}_N\phantom{_{1-2}}  & -  \chi \mathbf{I}_N\phantom{_{1-2}} \\
	   \chi \mathbf{I}_N &                      &                      &     \chi \mathbf{I}_N & -8 \chi \mathbf{I}_N & -\mathbf{A}_{M-2}                    &  8 \chi \mathbf{I}_N\phantom{_{1-2}} \\
	 8 \chi \mathbf{I}_N &   \chi \mathbf{I}_N  &                      &                       &    \chi \mathbf{I}_N & -8 \chi \mathbf{I}_N\phantom{_{1-2}} & -\mathbf{A}_{M-1}                    \\
\end{bmatrix*}
\end{equation}
with $\mathbf{A}_i = \mathbf{A}(s_i)\in\mathbb{R}^{N\times N}$ and $\mathbf{I}_{N}$ the identity matrix of appropriate size, arises from the temporal discretisation and has a banded structure with $2N\times 2N$ blocks on the corners arising from the periodic boundary conditions \cref{eq:standard-form-bc}. The system \cref{eq:bordered-system} is well conditioned by the additional row arising from the phase locking constraint, even when $\mathbf{D}$ is singular, i.e. on a periodic orbit. For the adjoint problems \cref{eq:adjoint-problem} and \cref{eq:homo-adjoint-problem}, where technically the term $\kappa\mathbf{c}(s)$ in \cref{eq:standard-form} is zero, we use the same approach and check that $\kappa$ is zero to machine accuracy.

We solve the bordered system \cref{eq:bordered-system} using specialised dense linear algebra techniques. The mixed block elimination method BEM is used \cite{Govaerts:1994dc}. The algorithm is backward stable regardless of the condition number of $\mathbf{D}$, if a backward stable solver is available for $\mathbf{D}$ and its transpose. We exploit the cyclic block tri-diagonal structure of $\mathbf{D}$ (actually penta-diagonal, but blocks are lumped together in groups of four for simplicity) and used the algorithm described in Reference \cite{Batista:2009gm} for solving the linear systems arising in the BEM algorithm. This uses at its core the Sherman-Morrison-Woodbury formula to handle the corner blocks. LU factorisation and solution of banded linear systems (the banded body of $\mathbf{D}$) with bandwidth $4N+1$ and $M$ rows is performed using \texttt{LAPACK} implementations of \texttt{gbtrf} and \texttt{gbtrs}, respectively. The vast majority of the computational time is spent in calls to these two functions, hence the overall method benefits from using efficient implementations. Note that the factorisation of the banded body of $\mathbf{D}$ can be reused for the solution of the homogeneous and inhomogeneous adjoint problems \cref{eq:adjoint-problem} and \cref{eq:homo-adjoint-problem}.

{Solving \cref{eq:standard-form} with the proposed finite difference approach becomes impractical for high-dimensional systems, such as, for instance, discretisations of Navier-Stokes fluid problems in three dimensional domains. In particular, the storage of the ``space-time'' trajectory $\hat{\mathbf{z}}$ can be problematic, but more importantly the construction and solution of problem \cref{eq:bordered-system} can be a formidable challenge. A better approach in this scenario, would be to solve the relevant BVPs using shooting techniques \cite{Ascher:1994ty,Zahr:2016bd}, greatly reducing the storage burden.}

\section{Graph-based algorithm to construct a set of unique orbits}\label{app:graph}
Different initial guesses might converge to solutions that define essentially the same orbit. In some instances, solutions might also describe more than one loop around the same orbit. From a numerical viewpoint, distinguishing these duplicates is not immediate. Here, we describe an efficient graph-based algorithm, scalable to hundreds of thousands orbits, to identify and remove such duplicates. 

\begin{figure}[htbp]
	\centering
	\includegraphics[width=1\textwidth]{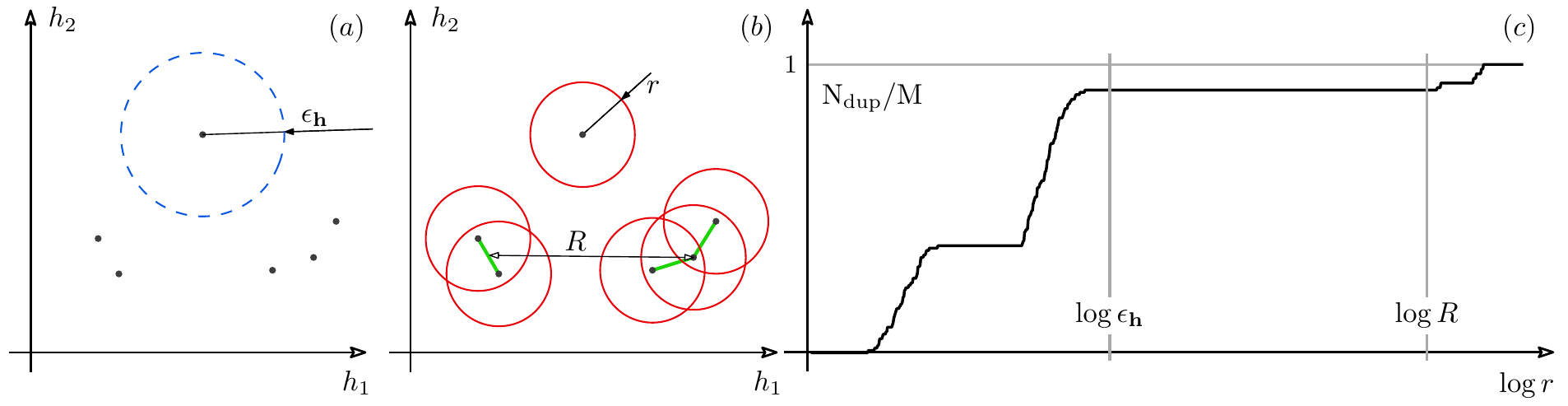}
	\caption{Panel (a): invariant features are calculated for each orbit up to an error estimate $\epsilon_\mathbf{h}$, resulting in a collection of points in a $N^\mathbf{h}$ dimensional space. Panel (b): a connectivity graph is built (green segments) by connecting points closer than a given distance $r$. The connected components of this graph are extracted, and the orbit with the shortest period is saved whilst duplicates are removed. Panel (c): typical effect of ball radius $r$ on the number of duplicates identified. }
	\label{fig:duplicate}
\end{figure}

\subsection*{Step 1: Calculation of orbit invariant features} A vector of invariant features $\mathbf{h} \in R^{N_\mathbf{h}}$ is first calculated for each orbit. The features can be thought of as quantities that characterise uniquely the location and structure of a periodic trajectory in state space and are invariant under time translation.  In this work, we used the orbit barycenter $\mathbf{h} = \langle \mathbf{x}^o(s)\rangle$ over the first six state variables for the Kuramoto-Sivashinsky problem and the three state variables for the Lorenz equations. If needed, the invariance can be extended to other symmetries, to avoid storing more than once the same orbit (this was done for the KSEq by taking the absolute value of the orbit barycenter). This procedure leads to a set of $M$ points $\mathcal{H} = \{\mathbf{h}^i\}_{i=1}^M$ in the ``feature space'', where $M$ is the number of converged solutions, \cref{fig:duplicate}-(a).  The error $\epsilon_\mathbf{h}$ on these invariants should be estimated. Here, we refined the temporal discretisation and used a Richardson extrapolation technique to estimate the error. Integrals were computed using a composite trapezoidal rule, which converges exponentially for periodic functions \cite{Trefethen:2014du}. This ensures that the error in the calculation of period averages was due to the truncation error of the nonlinear BVP, rather than to the numerical quadrature.

\subsection*{Step 2: Construction of connectivity graph using tree-based nearest neighbour search}
A connectivity graph is constructed by identifying the adjacency sets 
\begin{equation}
\mathcal{N}^i = \{\mathbf{h}^j \in \mathcal{H}\,:\, \mathrm{dist}(\mathbf{h}^j, \mathbf{h}^i) < r, i\neq j\},
\end{equation}
the sets of feature vectors that are contained in a ball of radius $r$ around $\mathbf{h}^i$, where $\mathrm{dist}(\cdot, \cdot)$ is an appropriate distance function, \cref{fig:duplicate}-(b). Using an adjacency set representation allows for efficient implementations of graph traversal algorithms. To build such sets we used an efficient space-partitioning data structure, a KD-tree, to perform the nearest neighbours searches. A KD-tree is a tree-based data structure that recursively partitions with a hyper-plane the (feature) space into two sub-partitions, each containing half the points (feature vectors), until all points are exhausted. Building a KD-tree out of $M$ points has $\mathcal{O}(M\log(M))$ complexity, whereas searching for known points in a given range around any point in the feature space has $\mathcal{O}(N_\mathbf{h}\cdot M^{1-1/N_\mathbf{h}})$ complexity. Hence, this structure enables significantly more efficient searches than a round-robin approach when $M\gg2^{N_\mathbf{h}}$.

\subsection*{Step 3: Identification of connected components}
We use a depth-first-search traversal algorithm to identify the connected components from the connectivity graph. Connected components identify clusters of solutions that have `similar' invariant features and thus describe the same UPO. This operation has, in total, $\mathcal{O}(M)$ complexity. For each connected component in the graph, the orbit with the shortest period is saved while the all others are tagged as duplicates and removed. This results in a set of unique orbits.

\subsection*{Step 4: Effect of ball radius $r$} 
Because the distribution of solutions in a high-dimensional feature space might be hard to visualise, we study the effect of the ball radius $r$ on the number of duplicates $N_{dup}$ found. \cref{fig:duplicate}-(c) shows the typical behaviour of the algorithm, applied to a large dataset of UPOs found by converging initial guesses from a long chaotic realisation of the KSEq. For $r>R$, roughly the minimum distance between the barycenters of the connected components, a large number of UPOs are identified incorrectly as duplicated, because individual clusters that represent different UPOs are connected together. For $r <\epsilon_\mathbf{h}$, nearby solutions approximating the same orbit form distinct clusters and are thus not tagged as duplicates. Any choice of  $r$ in the interval $\epsilon_{\mathbf{h}} \ll r \ll R$ will identify duplicates correctly. This shows that a sensible choice of $\epsilon_{\mathbf{h}}$ is possible an order of magnitude lower than $R$.

\subsection*{Further remarks}
In practice, the KD-tree based nearest neighbours search is significantly more efficient when $N_\mathbf{h} < \log_2(M)$. This limits the number of invariant features that can be used. In spatially extended systems with many degrees of freedom, good invariant features could be chosen by projecting UPOs onto a low-dimensional linear subspace obtained, for instance, by the span of the first leading Proper Orthogonal Decomposition vectors. The data transformation will ensure maximal efficiency, because the splitting hyper-planes in the construction of the KD-tree will be orthogonal to axis of inertia of the data points. 

\section{Sensitivity of long UPOs}\label{app:long-upo-lorenz}
We show, empirically, that the adjoint sensitivity calculations using the methods discussed in  \cref{app:upo-search-algo} and \cref{app:appendix-numerical-solution} are well behaved regardless of the stability characteristics and length of the UPO.  {To} this end we show convergence results for a rather long UPOs at $\rho=28$, with period $T \approx 1080$ and symbol sequence length 1439, too long to be reported here. An initial guess for the search is obtained by taking a point on the attractor, after transients have decayed, integrating the Lorenz equations for a thousand time units and then finding the nearest recurrence within the next hundred time units. Such long UPOs can be easily found using the approach described in \cref{app:upo-search-algo}, because the instability associated to classical shooting methods is completely avoided by the particular discretisation adopted \cite{Ascher:1994ty}.
 
\begin{table}
	\small
	\centering
	\caption{Convergence of UPO characteristics and their sensitivity at $\rho=28$. Converged digits are underlined.}
	\label{tab:long}
	\begin{tabular}{crcccc}
	 $\Delta t$ & $M$     & $T$           & $T_{\mathrm{d}\rho}$ & $\langle z \rangle $ & $\langle z \rangle_{\mathrm{d}\rho}$ \\
	 \hline
	 \hline
	 0.020000 &   54026 & \underline{1080}.4469534509 & \underline{-26.12}39369819 & \underline{23.5}498909445 &  \underline{1.016}6218037 \\
	 0.010000 &  108052 & \underline{1080.51}31786996 & \underline{-26.124}7514912 & \underline{23.551}7275703 &  \underline{1.0165}702500 \\
	 0.005000 &  216104 & \underline{1080.517}3671241 & \underline{-26.12480}30601 & \underline{23.5518}442606 &  \underline{1.016566}9625 \\
	 0.002500 &  432208 & \underline{1080.5176}296848 & \underline{-26.124806}2847 & \underline{23.55185}15967 &  \underline{1.0165667}556 \\
	 0.001250 &  864416 & \underline{1080.51764}61071 & \underline{-26.1248064}862 & \underline{23.5518520}559 &  \underline{1.01656674}26 \\
	 \hline
	\end{tabular}
\end{table}

\begin{figure}[htbp]
	\centering
	\includegraphics[width=0.94\textwidth]{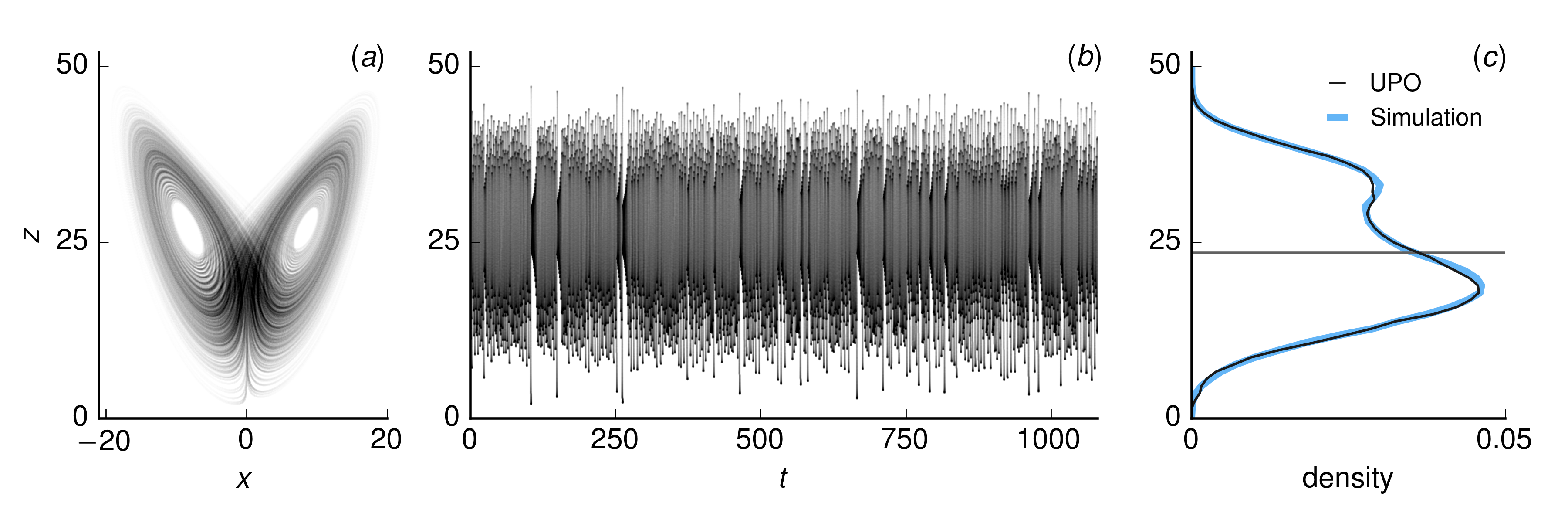}
	\caption{A long UPO of the Lorenz equations. Panel $(a)$, projection on the $(x, z)$ plane; panel $(b)$, time history of $z(t)$; panel $(c)$, normalised histogram of the $z$ variable over 100 bins from 0 to 50, for the UPO and for a long chaotic trajectory (10000 time units).} 
	\label{fig:long-orbit}
\end{figure}

The rows of \cref{tab:long} show how the same quantities reported in \cref{tab:lorenz} vary as the temporal discretisation is refined. Refinement is performed by doubling the number of grid points ($M$) and using a linear prolongation algorithm to interpolate the solution on the finer grid. Newton-Raphson iterations are then run to convergence with less tight tolerances due to the length of the orbit ($\alpha=11$ as discussed in \cref{app:appendix-numerical-solution}). At each refinement iteration approximately one digit of accuracy is gained, due to the fourth order accurate scheme \cref{eq:discretisation-fourth-order}. This UPO is shown in \cref{fig:long-orbit}. What is remarkable and intriguing is that the temporal dynamics of this orbit is indistinguishable from a chaotic trajectory, as they are its statistics (e.g. the density of $z$, in \cref{fig:long-orbit}-(c)). The difference, however, is that unlike for similarly long chaotic trajectories, its sensitivity with respect to parameter perturbations can be calculated exactly.

\bibliographystyle{siamplain}
\bibliography{../../library}

\end{document}